\documentclass[12pt,titlepage]{article}

\font\grande=cmr9.5 scaled \magstep4
\font\medio=cmr9.5 scaled \magstep2
\outer\def\beginsection#1\par{\medbreak\bigskip
      \message{#1}\leftline{\bf#1}\nobreak\medskip
\vskip-\parskip
      \noindent}
\usepackage{graphicx} 
\begin{document}
\bibliographystyle {unsrt}

\titlepage

\begin{flushright}
CERN-PH-TH/2010-242
\end{flushright}

\vspace{1cm}
\begin{center}
{\grande Hanbury Brown-Twiss interferometry}\\
\vspace{5mm}
{\grande and second-order correlations of inflaton quanta}\\
\vspace{1cm}
 Massimo Giovannini 
 \footnote{Electronic address: massimo.giovannini@cern.ch} \\
\vspace{1cm}
{{\sl Department of Physics, 
Theory Division, CERN, 1211 Geneva 23, Switzerland }}\\
\vspace{0.5cm}
{{\sl INFN, Section of Milan-Bicocca, 20126 Milan, Italy}}
\vspace*{1cm}
\end{center}

\vskip 0.3cm
\centerline{\medio  Abstract}
\vskip 0.1cm
The quantum theory of optical coherence is applied to the scrutiny of the statistical properties of 
the relic inflaton quanta. After adapting the description of the quantized 
scalar and tensor modes of the geometry to the analysis of intensity 
correlations, the normalized degrees of first-order and second-order coherence are computed in the concordance paradigm and are shown to encode faithfully the statistical properties of the initial quantum state.  The strongly bunched curvature phonons are not only super-Poissonian but also super-chaotic.  Testable inequalities are derived in the limit of large angular scales and can be physically interpreted in the light of the tenets of Hanbury Brown-Twiss interferometry.   The quantum mechanical results are compared and contrasted with different situations including the one where intensity correlations are the result of a classical stochastic process. The survival of second-order correlations (not necessarily related to the purity of the initial quantum state) is addressed by defining a generalized ensemble where super-Poissonian statistics is an intrinsic property of the density matrix and turns out to be associated with finite volume effects which are expected to vanish in the thermodynamic limit.  
\noindent

\vspace{5mm}
\vfill
\newpage
\renewcommand{\theequation}{1.\arabic{equation}}
\setcounter{equation}{0}
\section{Formulation of the problem}
\label{sec1}
The current data accounting for various properties of the Cosmic Microwave Background (CMB) 
anisotropies and polarization are all consistent with the presence, prior to matter-radiation equality, of a quasi-flat 
spectrum of curvature perturbations whose origin is customarily attributed to an (early) 
inflationary stage of expansion occurring when the Hubble rate was, roughly, of the 
order of one millionth of the Planck energy scale (see Refs. \cite{WMAP7a,WMAP7b,ACBAR,QUAD} and,  for a more 
comprehensive perspective on large-scale data, also Ref. \cite{wein}).  When CMB data are combined with other data sets (such as the large-scale structure data \cite{LSS1,LSS2} and the supernova data \cite{SNN1,SNN2}) the typical parameters describing the large-scale curvature 
modes are slightly (but not crucially) modified.  The agreement between the parameter 
determinations obtained by combining different 
data sets seems to reach the level of the per mill. This apparent (statistical) accuracy can be reached within the 
simplest scenario, conventionally dubbed $\Lambda$CDM where $\Lambda$ stands for the dark energy component 
and CDM for the cold dark matter component. The quest for statistical accuracy should not forbid to think, in broader terms, 
to the very nature of the pre-inflationary initial conditions whose precise nature is all but well established \cite{wein} and anyway 
not explained in the framework of the $\Lambda$CDM paradigm.

Can we establish, independently of the CMB data sets (see e.g. Refs. \cite{WMAP7a,WMAP7b,QUAD,ACBAR}), 
the duration of the inflationary stage of expansion\footnote{In the $\Lambda$CDM paradigm, the total curvature of the Universe receives, at the present time, a leading contribution from the extrinsic curvature and a subleading 
contribution from the intrinsic (spatial) curvature. The role of inflation is, in this context, to 
make the ratio between intrinsic and extrinsic curvature sufficiently minute at the onset of the radiation 
dominated epoch so that it can easily be of order $1$ today (during a decelerated stage of expansion 
the ratio between intrinsic and extrinsic curvature is actually increasing). The duration 
of the inflationary phase required to solve the latter goal represents one of the ways of pinning 
down the minimal number of inflationary efolds.}? 
Not really: inflation could have been very long or it could have been just minimal (i.e. with approximate 
duration between $63$ and $65$ e-folds by assuming the largest value of the slow-roll parameter 
compatible with the current upper limits set by the WMAP data \cite{WMAP7a,WMAP7b}).
Is it known which were the initial conditions of the scalar and tensor modes 
of the geometry around the onset of the inflationary dynamics? Not really: if inflation lasted much more than the required 65 e-folds probably the only sound initial state for the inflaton quanta (and within the logic of inflationary models) was the vacuum.  Conversely, if the duration of inflation was just minimal (or close to minimal) then different kinds of quantum mechanical initial states could play a decisive role and their associated energy density can only be constrained by back-reaction considerations \cite{mg1}. 

The pair of questions contained in the previous paragraph are usually answered within two opposite points of view. 
Within the first  set of hypotheses, the duration of the inflationary phase is determined by the aim of reproducing 
some structure in the temperature or polarization angular power spectra: in this case the duration 
of the inflationary phase is correlated (by construction) with the spectral  behaviour of the initial 
data for the scalar and tensor modes of the geometry. The second (often tacit) assumption is that 
the duration of inflation is much larger than $60$ efolds: in this way quantum mechanical initial conditions are better justified but 
the initial state does not have specific observable consequences. In different terms, the present measurements 
on the large-scale temperature and polarization anisotropies are only able to probe the power spectrum of curvature (adiabatic)
fluctuations possibly present prior to matter-radiation equality over typical wavelengths much larger 
than the Hubble radius at the corresponding epoch. The increase of the number of inflatons or the addition 
of supplementary (non adiabatic) components in the initial conditions \cite{nad1,nad2,nad3} automatically increases the number of parameters by making the model less predictive even if, potentially, more sound. 

\begin{figure}[!ht]
\centering
\includegraphics[height=6cm]{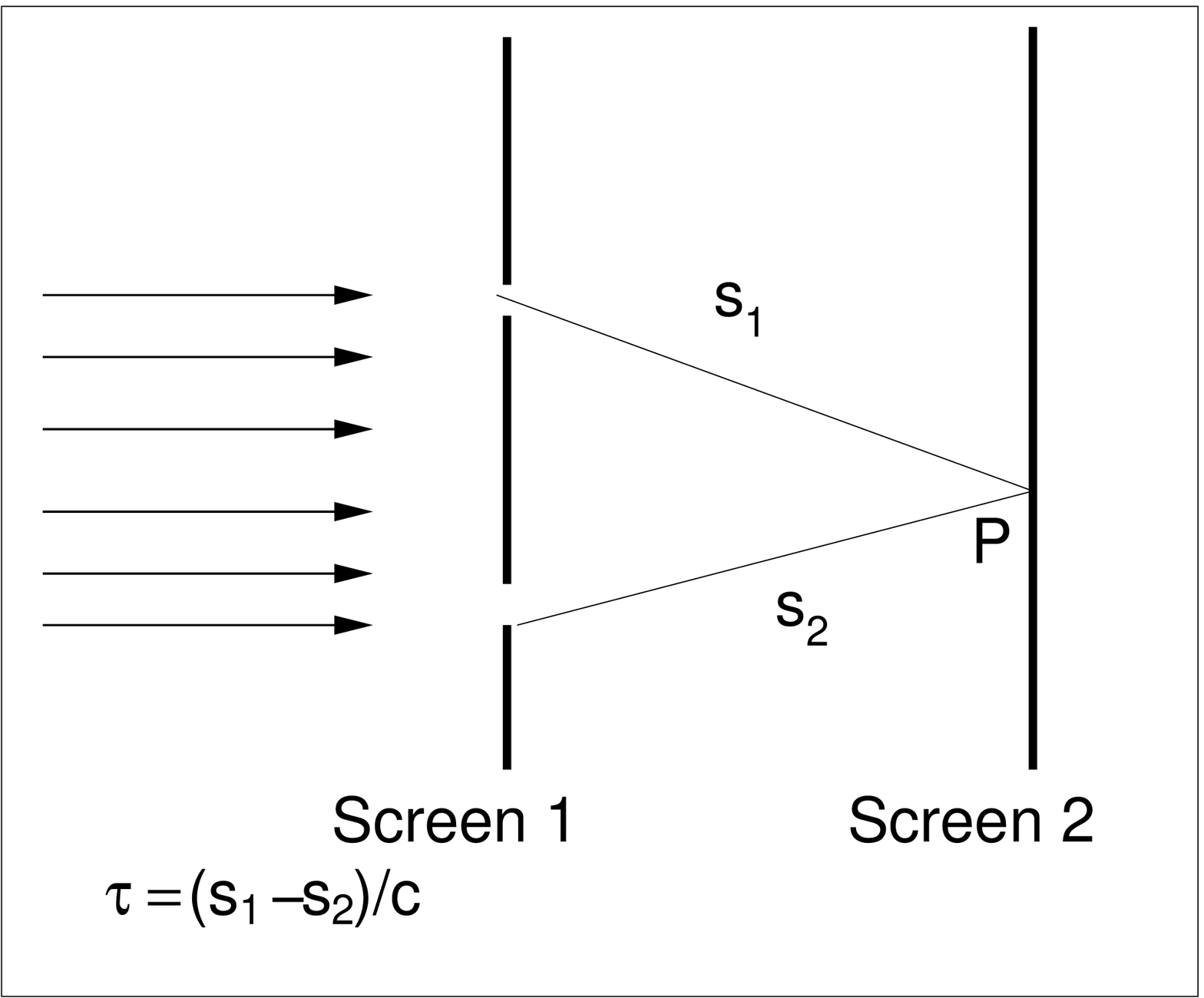}
\includegraphics[height=6cm]{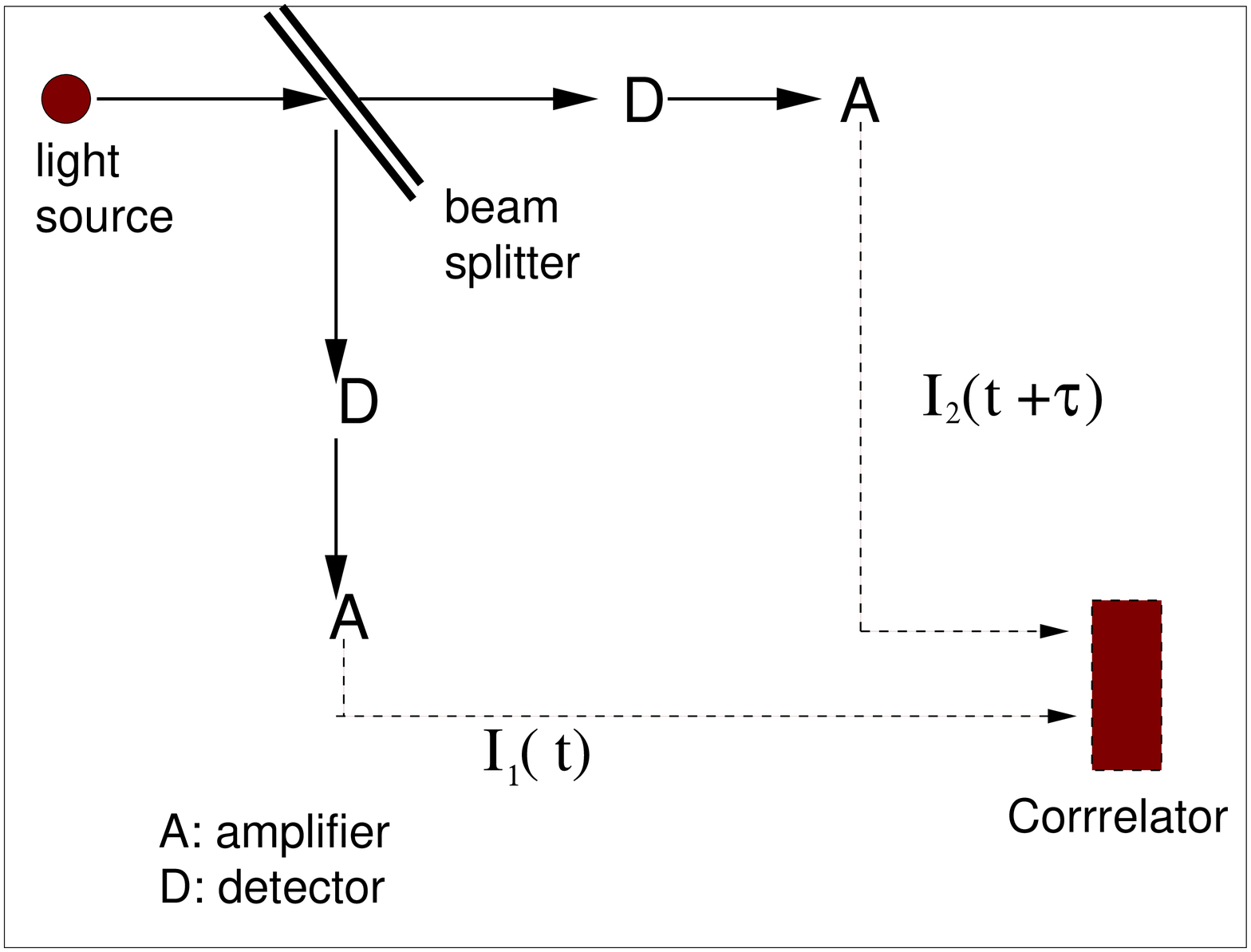}
\caption[a]{The Young (plot on the left) and Hanbury Brown-Twiss (plot on the right) experiments are compared; $\tau$ denotes the time 
delay (the speed of light $c$ has been restored while natural units 
$\hbar = c =1$ will be used throughout).}
\label{figure1A}      
\end{figure}

The aim of the present paper is less pretentious than the two extreme 
approaches mentioned in the previous paragraph: instead of arguing (on a purely theoretical 
basis) how long inflation must have been, it seems plausible to scrutinize wether it is possible, at least 
in principle, to determine the statistical properties of the initial state of relic inflaton quanta. 
In this respect it is both plausible and useful to draw a physical analogy with a similar class of problems arising 
in the quantum optical treatment of the fluctuations of visible light (see \cite{sudarshan,loudon,mandel}  for three classic 
treatises covering, with different emphasis, all the theoretical tools which will also play a role in the forthcoming 
considerations).  Consider, for sake of concreteness, a scalar quantum field $\hat{V}(\vec{x},\tau)$ where $\vec{x}$ denotes the 
spatial coordinate and $\tau$ the time variable; $\hat{V}(\vec{x},\tau)$ might denote the quantum field describing either relic phonons or the single polarization of a graviton; in the following discussion, however, the scalar field $\hat{V}(\vec{x},\tau)$ denotes the single polarization of an electromagnetic wave in the visible frequencies as sometimes done in quantum optics 
\cite{sudarshan}.  The correlation function\footnote{The averages $\langle\, ...\,\rangle$ can denote either ensemble (statistical) or quantum averages depending on the nature 
of the source and also upon the preferred physical description. In the present paper
the amplitude and the intensities are always related to quantum fields unless 
explicitly stated (see, in particular, section \ref{sec6}). }:
\begin{equation} 
G^{(1)}(\vec{x}, \vec{y}; \tau_{1},\tau_{2}) = \langle \hat{V}(\vec{x},\tau_{1}) \hat{V}(\vec{y},\tau_{2})\rangle
\label{EQ1}
\end{equation} 
can be probed, in quantum optics, by the Young two-slits 
experiment which is sensitive to the interference of the amplitudes of the  radiation field 
\cite{mandel} (see Fig. \ref{figure1A}). Young interferometry is not able, by itself, to provide information on the statistical properties of the quantum state of the radiation field since various states with diverse physical properties (such as laser light and chaotic light) 
lead to comparable degrees of first-order coherence.  The normalized counterpart of Eq. (\ref{EQ1}) can be written as
\begin{equation} 
g^{(1)}(\vec{x}, \vec{y}; \tau_{1},\tau_{2}) = \frac{\langle \hat{V}(\vec{x},\tau_{1}) \hat{V}(\vec{y},\tau_{2})\rangle}{\sqrt{\langle| \hat{V}(\vec{x},\tau_{1})|^2 \rangle \langle| \hat{V}(\vec{y},\tau_{2})|^2\rangle}},
\label{EQ1A}
\end{equation}
defines the degree of first-order coherence. In Young interferometry the 
light field can be classified depending upon the value of $g^{(1)}(\vec{x}, \vec{y}; \tau_{1},\tau_{2})$: 
\begin{itemize}
\item{} the light is first-order coherent provided $g^{(1)}(\vec{x}, \vec{y}; \tau_{1},\tau_{2})=1$;
\item{} the light is partially coherent is $0< g^{(1)}(\vec{x}, \vec{y}; \tau_{1},\tau_{2}) <1$;
\item{} the light is incoherent if $g^{(1)}(\vec{x}, \vec{y}; \tau_{1},\tau_{2}) =0$.
\end{itemize}
In the Young two-slit experiment the electric fields emerging from the two pinholes produce the interference fringes on the second screen. The maximal (total) intensity on the second screen can be written as 
${\mathcal I}_{\mathrm{max}} = {\mathcal I}_{1} + {\mathcal I}_{2} + 2 \sqrt{{\mathcal I}_{1} {\mathcal I}_{2}} g^{(1)}(\tau)$ 
while the minimal  (total) intensity on the second screen can be written as 
${\mathcal I}_{\mathrm{min}} = {\mathcal I}_{1} + {\mathcal I}_{2} - 2 \sqrt{{\mathcal I}_{1} {\mathcal I}_{2}} g^{(1)}(\tau)$ 
where ${\mathcal I}_{1}$ and ${\mathcal I}_{2}$ denote the intensity of the radiation field in each of the two 
pinholes. The visibility, i.e. $({\mathcal I}_{\mathrm{max}} - {\mathcal I}_{\mathrm{min}})/({\mathcal I}_{\mathrm{max}} + {\mathcal I}_{\mathrm{min}})$
coincides exactly with $g^{(1)}(\tau)$ in the case  ${\mathcal I}_{1} = {\mathcal I}_{2}$. If $g^{(1)}(\tau) =1$ 
the visibility is maximized and, as discussed before, the light is said to be first-order coherent. 

Until the mid fifties, Eq. (\ref{EQ1}) has been used to define the coherence of the radiation field: a field was said to be coherent when the interference fringes are maximized in Young-type (two-slit) correlation experiment of the type of the one
reported in the left plot of Fig. \ref{figure1A}.  The applications of the Hanbury Brown-Twiss (HBT) effect (first to stellar interferometry \cite{HBT1} and then 
more specifically to quantum optics \cite{HBT2}) demanded the accurate 
study of not only the correlations between field strengths (as defined in Eq. (\ref{EQ1})) but also 
the analysis of the correlations between the intensities of the radiation field, i.e. 
\begin{equation}
G^{(2)}(\vec{x}, \vec{y}; \tau_{1}, \tau_{2}) = \langle \hat{{\mathcal I}}(\vec{x},\tau_{1}) \hat{{\mathcal I}}(\vec{y},\tau_{2}) \rangle 
\label{EQ2}
\end{equation}
where now $\hat{{\mathcal I}}(\vec{x},\tau)  =  \hat{V}^2(\vec{x},\tau)$ is the intensity of the radiation field. In the realistic case 
when the radiation field is described by the electric field the intensity will simply be the squared modulus 
of the electric field. The physical implications of HBT interferometry in general and of Eq. (\ref{EQ2}) in particular 
have been important for many areas of physics ranging from stellar astronomy \cite{HBT1} and quantum optics
\cite{sudarshan,loudon,mandel},  to pion interferometry  \cite{podgo1,podgo2,cocconi} and subatomic physics (see \cite{rev1,rev2} for two comprehensive reviews). In subatomic physics HBT interferometry has been used to determine the hadron fireball dimensions \cite{cocconi} which is related to the linear size of the interaction region in proton-proton collisions.  
In Fig. \ref{figure1A} Young interferometry (plot on the left) is schematically compared to HBT interferometry (plot at the right).  
In short the difference between the two experiments resides 
in the order of the correlation. In the case of Young interferometry (plot 
on the left in Fig. \ref{figure1A}) the fringes on the second screen 
arise as the result of the interference electric field amplitudes. In the case of HBT 
interferometry the intensities of the radiation field 
are measured at the correlator (see right plot in Fig. \ref{figure1A}). Since 
the intensities are quadratic in the amplitude of the electric fields, HBT 
involves the study of second-order correlation effects. Furthermore, since in quantum theory the intensities 
of the radiation field are quantized, HBT correlations represented some of the first evidence of quantum 
effects in the description of optical fields \cite{sudarshan} as neatly expressed by Glauber \cite{glauber1,glauber2} 
(see, in particular, section IV of the first paper quoted in Ref. \cite{glauber2}).

The Glauber theory of optical coherence \cite{glauber1,glauber2} 
generalizes the concept of first-order coherence to higher orders and, in particular, to second-order. The second-order 
correlator defined in Eq. (\ref{EQ2}) can be written in its normalized form in full analogy with Eq. (\ref{EQ1A}). 
In the case when the intensities are purely classical  stochastic variables the degree of second-order coherence can be written as 
\begin{equation}
\overline{g}^{(2)}(\vec{x},\vec{y}; \tau_{1},\tau_{2}) = \frac{\langle{\mathcal I}(\vec{x},\tau_{1}) \,{\mathcal I}(\vec{y},\tau_{2})\rangle}{\langle 
{\mathcal I}(\vec{x},\tau_{1})\rangle \,\langle
{\mathcal I}(\vec{y},\tau_{2})\rangle}.
\label{EQ3a}
\end{equation}
It is also common to define a slightly different normalized correlator 
(see, e.g. \cite{rev1})
\begin{equation}
R(\vec{x},\vec{y},\tau) = \frac{\langle{\mathcal I}(\vec{x},\tau) \,{\mathcal I}(\vec{y},\tau)\rangle}{\langle 
{\mathcal I}(\vec{x},\tau)\rangle \,\langle
{\mathcal I}(\vec{y},\tau)\rangle} -1.
\label{EQ3b}
\end{equation}
If the intensities are constructed from an appropriate field operator, Eq. (\ref{EQ3a}) is usually written, in a quantum 
optical context, as
\begin{equation}
\overline{g}^{(2)}(\vec{x},\vec{y}; \tau_{1},\tau_{2}) = \frac{\langle :\hat{{\mathcal I}}(\vec{x},\tau_{1}) \, \hat{{\mathcal I}}(\vec{y},\tau_{2}): \rangle}{\langle: 
\hat{{\mathcal I}}(\vec{x},\tau_{1}):\rangle \,\langle:
\hat{{\mathcal I}}(\vec{y},\tau_{2}):\rangle},
\label{EQ3}
\end{equation}
where the colon makes explicit the normal ordering of the operators. In quantum optics
it is natural to impose the normal ordering in the correlators since the detection 
of light quanta (i.e. in the optical range of frequencies) occurs by detecting a current 
induced by the absorption of a photon \cite{glauber1,glauber2}. In Fig. \ref{figure1A} (plot at the right) the 
basic logic of the HBT experiment is schematically illustrated: the electric field is first split into two components 
through the beam splitter, then it is time-delayed and finally recombined at the correlator. The HBT setup 
provides therefore an operational definition for correlating the {\em intensities} of the radiation field. Conversely Young 
interferometry only probes the correlations between the {\em amplitudes} of the radiation field.  According to the Glauber theory of optical coherence, 
the radiation field is said to be {\em first-order coherent} if 
$g^{(1)}(\vec{x}, \vec{y}; \tau_{1},\tau_{2}) =1$; the radiation field is said to be 
second-order coherent if $g^{(2)}(\vec{x}, \vec{y}; \tau_{1},\tau_{2})=1$.
As we shall see in a moment, the coherent states of the radiation 
field are both first-order and second-order coherent.

\begin{figure}[!ht]
\centering
\includegraphics[height=6cm]{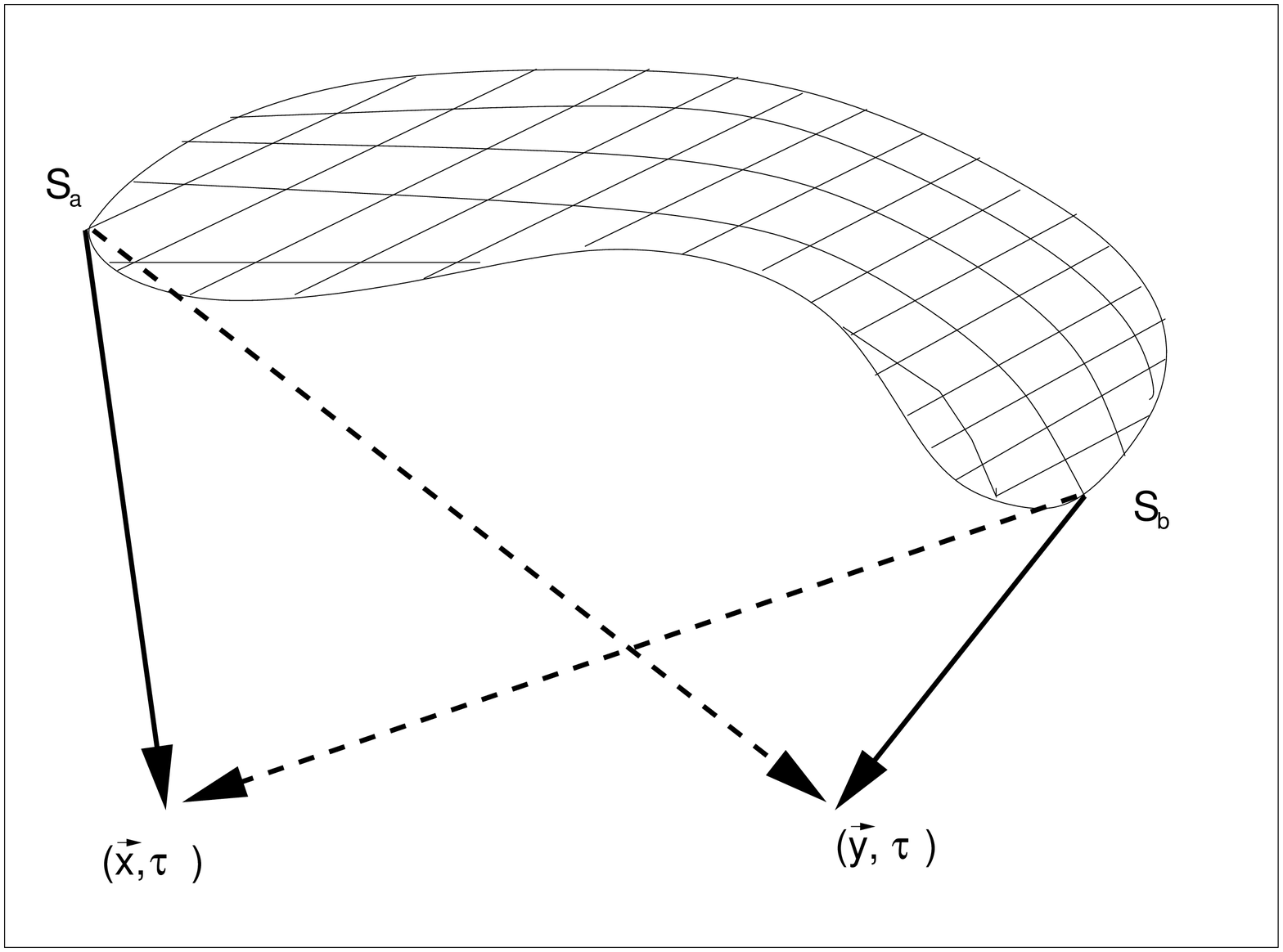}
\includegraphics[height=6cm]{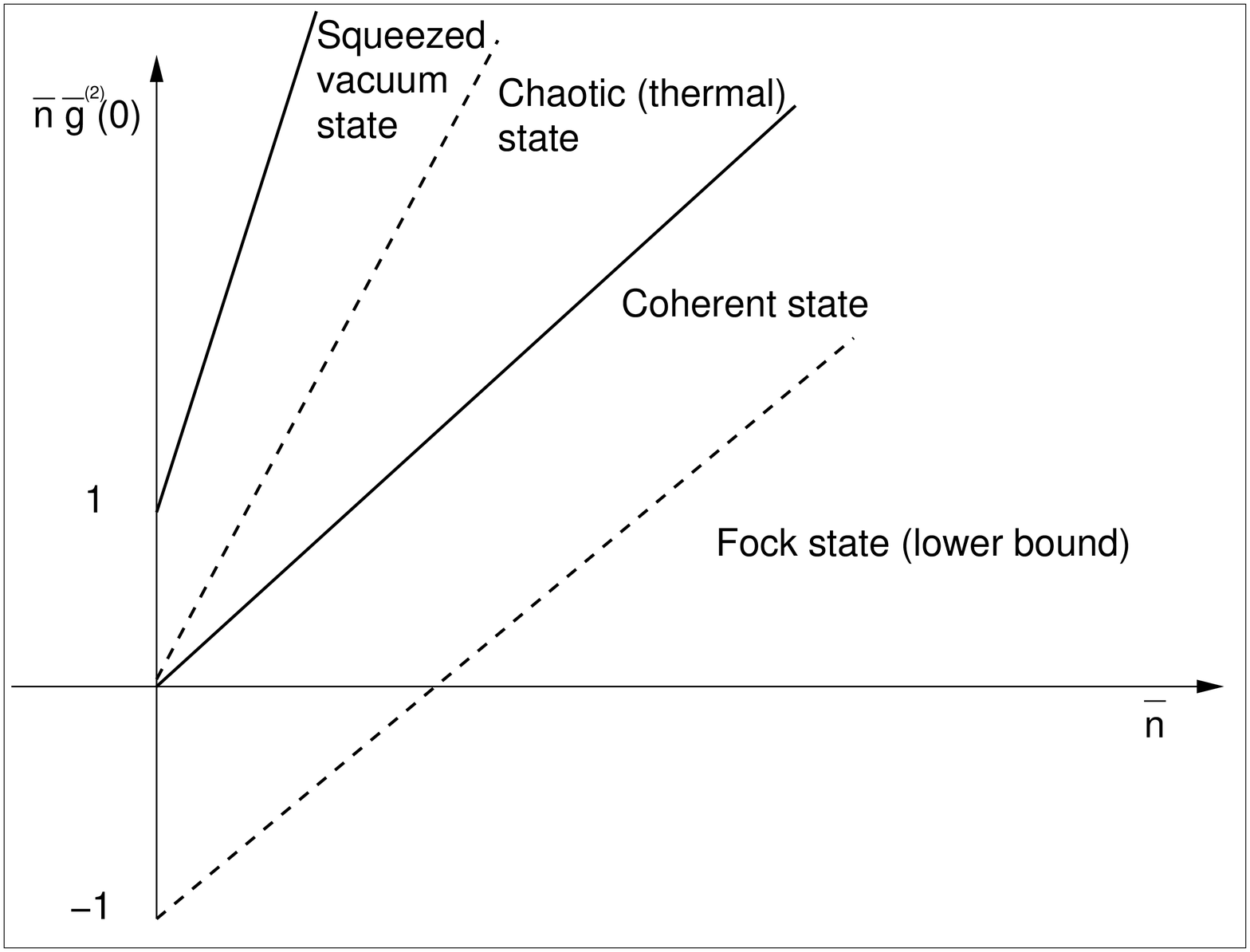}
\caption[a]{In the plot on the left the physical idea of  interfering intensities is schematically illustrated. In the plot at the right 
the different values of the intercept $\overline{n}\, \overline{g}^{2}(0)$ are reported for different quantum states 
as a function of the average multiplicity of each quantum state.}
\label{figure1}      
\end{figure}

HBT interferometry encodes two complementary pieces of information characterizing the source, i.e. 
\begin{itemize}
\item{} the linear (or angular) size of the emitting (hyper)surface;
\item{} the statistical properties of the emitting quanta (photons, pions, phonons, gravitons).
\end{itemize} 
The first aspect is illustrated in the left plot of Fig. \ref{figure1} where the correlations of the 
intensities of the radiation field ${\mathcal I}(\vec{x},\tau_{1})$ and ${\mathcal I}(\vec{y},\tau_{2})$ 
are depicted in the situation where $\tau_{1} = \tau_{2} = \tau$. 
In Eqs. (\ref{EQ3a}) and (\ref{EQ3b}) $\langle{\mathcal I}(\vec{x},\tau) \,{\mathcal I}(\vec{y},\tau)\rangle$ denotes the intensities measured both in $\vec{x}$ and in $\vec{y}$
while  $\langle {\mathcal I}(\vec{x},\tau)\rangle$ and $\langle {\mathcal I}(\vec{y},\tau)\rangle$
denote the intensities measured separately in the two points. This definition is schematically illustrated in the left plot of Fig. \ref{figure1}. 
In a quantum mechanical perspective Eqs. (\ref{EQ3b})--(\ref{EQ3}) 
imply that the normalized degree of second-order coherence 
can be measured by counting the photons. More specifically, supposing 
that $\tau_{1} = \tau_{2}$ the number of particles 
observed simultaneously in $\vec{x}$ and $\vec{y}$    
can be divided by the product of the number of counts observed separately 
in $\vec{x}$ and $\vec{y}$ (see also Fig. \ref{figure1}). The same reasoning 
holds, of course, also in the case of pions \cite{cocconi} as well as in the 
case of other particles obeying the Bose-Einstein statistics 
such as relic phonons and relic gravitons. In this sense HBT interferometry 
is deeply connected to what is known, in high-energy physics, as the 
study of Bose-Einstein correlations \cite{cocconi,rev1}.

In the case of the relic phonons and gravitons the normal ordering of the correlators is not specifically justified (even if it is technically useful, as we shall see). One of the purposes of the present study is to give the correct quantum mechanical definition of the degree of second-order coherence in the case of the relic phonons and of the relic gravitons in strongly correlated quantum states with large occupation numbers per Fourier mode.

If the detection of the intensity occurs at the same spatial location (i.e. $\vec{x} = \vec{y}$) 
the degree of second-order coherence will depend upon the time difference $\tau = \tau_{1} - \tau_{2}$
with a bell-like shape. For $\tau \to 0$  (zero time-delay limit)  the degree of second-order coherence reaches a specific value 
which depends upon the statistical properties of the source and which will be discussed in a moment. When 
$g^{(2)}(\tau) \to 1$ in the limit $\tau \gg 1$, the  width of $g(\tau)$  estimates the coherence time of the source. A similar discussion 
can be conducted for the degree of space-time coherence and allows, in pion physics, an approximate determination 
of the hadronic fireball dimensions \cite{cocconi} (see also \cite{rev1,rev2}).

If photons are detected by photoelectric counting \cite{loudon},  Eq. (\ref{EQ3}) can be written, for  a single of the radiation field and in the limit $\tau \to 0$ (i.e. zero time-delay limit) as 
\begin{equation}
\overline{g}^{(2)}(0) = \frac{\langle \hat{a}^{\dagger}\, \hat{a}^{\dagger} \, \hat{a} \, \hat{a} \rangle}{\langle \hat{a}^{\dagger} \hat{a} \rangle^2 } = \
\frac{D^2 - \langle \hat{N} \rangle}{ \langle \hat{N}\rangle^2} + 1 ,\qquad D^2 = \langle \hat{N}^2 \rangle 
- \langle \hat{N} \rangle^2,
\label{EQ4}
\end{equation}
where $\hat{a}$ and $\hat{a}^{\dagger}$ obey the usual Heisenberg-Weyl algebra $[\hat{a},\hat{a}^{\dagger}] =1$ and $D^2$ denotes the variance.
To pass from the first equality in Eq. (\ref{EQ4}) to the second expression involving $D^2$ it must be noted that, using the 
commutation relations, $\langle \hat{a}^{\dagger} \hat{a}^{\dagger} \hat{a} \hat{a} \rangle = \langle \hat{N}^2 \rangle - \langle \hat{N} \rangle$ 
where $\hat{N} = \hat{a}^{\dagger} \hat{a}$. Using now the definition of $D^2$ we also have that $\langle \hat{a}^{\dagger} \hat{a}^{\dagger} \hat{a} \hat{a} \rangle
= [ D^2 -  \langle \hat{N} \rangle +  \langle \hat{N}\rangle^2]$. The second equality in Eq. (\ref{EQ4}) is finally proven by appreciating
 that, in the definition of $\overline{g}^{(2)}(0)$, the 
term $\langle \hat{a}^{\dagger}\, \hat{a}^{\dagger} \, \hat{a} \, \hat{a} \rangle$ is divided by $\langle \hat{a}^{\dagger} \hat{a} \rangle^2 $.

 From Eq. (\ref{EQ4}) it is clear that different quantum states will lead to different values of $g^{(2)}(0)$. For practical applications it is 
useful to define the so-called Mandel parameter (related to the zero-time-delayed correlator of Eq. (\ref{EQ3b})) whose expression
is 
\begin{equation}
{\mathcal Q} = \langle \hat{N} \rangle [ \overline{g}^{(2)}(0) -1] = \frac{D^2}{\langle \hat{N} \rangle} -1.
\label{EQ5}
\end{equation} 
In the case of a coherent state \cite{glauber1,glauber2}
\begin{equation}
D^2 = \langle \hat{N} \rangle, \qquad \overline{g}^{(2)}(0) =1,\qquad  \mathrm{and}\qquad {\mathcal Q}=0;
\label{EQ5a}
\end{equation}
and the radiation field is said to be second-order coherent. If the quantum state coincides 
with a Fock state containing $n$ particles (i.e. $| n \rangle$), the Mandel parameter equals $-1$.  The (single mode) 
Fock states defines the lower limit of the degree of second-order coherence i.e.   
${\mathcal Q} \geq -1$  and $\overline{g}^{(2)}(0) \geq 1 - 1/\langle \hat{N} \rangle$ (the equality is reached 
exactly in the case of a Fock state). 
Chaotic (e.g. white) light has the property of leading to a degree of second-order coherence double than in the case of a coherent  state, i.e. $\overline{g}^{(2)}(0) = 2$ which is a direct consequence of the (single  mode) density matrix for a thermal state (see 
e.g. \cite{sudarshan}).  The quantum 
mechanical correlations are then reflected in the degree of second-order coherence and, ultimately, in the 
magnitude and sign of $\overline{g}^{(2)}(0)$.  In Fig. \ref{figure1} 
 $ \langle \hat{N} \rangle \overline{g}^{2}(0)$ is illustrated for different quantum states as a function 
 of the average multiplicity $\overline{n} =  \langle \hat{N} \rangle$ of each state. To distinguish 
 graphically the different states it is practical to plot $\overline{n} \,\overline{g}^{(2)}(0)$ as a function of $\overline{n}$ 
 (rather than $\overline{g}^{(2)}(0)$ itself as a function of $\overline{n}$). Chaotic light is an example of bunched 
 quantum state (i.e. $\overline{g}^{(2)}(0) > 1$ implying more degree of second-order coherence 
 than in the case of a coherent state). Fock states are instead antibunched (i.e. $\overline{g}^{(2)}(0) <1$) implying 
 a degree of second-order coherence smaller than in the case of a coherent state. Experimentally the zero time-delay limit is 
 justified because the counting of photons (or pions) is made for typical times 
 smaller than the coherence time of the source. In this sense bunched particles  tend to arrive at the photodetector
 more simultaneously  than their antibunched counterpart. The concept of bunching will be relevant for a complete understanding of the physical properties of curvature phonons (see, e. g. sections \ref{sec2} and \ref{sec6}). 
 
In this paper it will be argued that the tenets of the quantum theory of optical coherence can be used to fully  characterize the correlation properties of cosmological perturbations. To explore different sets of initial conditions in conventional 
inflationary models the idea has been often to play either with a pre-inflationary phase of limited duration 
or to assign, in fully equivalent terms, an initial state on a given space-like hypersurface (see e.g. \cite{mg1,mg2,mg3} and 
references therein). Different initial states result in large-scale modifications of the power spectrum which can be 
used either to suppress or to increase the power at large scales \cite{mg3} (see also \cite{mg3a,mg3b}). 
Still, as it will be shown, different  initial state lead to the same degree of first-order coherence.
Second-order interference effects (and second-order coherence) provide a framework where the  statistical properties of the initial state can be classified and understood. The final aim of the approach pursued in this paper would be to reconstruct, 
by direct analysis of temperature and polarization correlations, the analog of the Mandel parameter and the degree of second-order 
coherence of the pre-decoupling initial conditions.

To pursue the program briefly outlined in the previous paragraph the first step is to apply and translate the theory of
(quantum) optical coherence to the case of relic inflaton quanta (i.e. relic gravitons and relic curvature phonons).  
The layout of the paper is therefore the following. Section \ref{sec2} contains a quantum mechanical premise where the second-order 
correlations are examined for a single degree of freedom (e.g. a mode of a cavity) but for 
quantum states whose statistical properties are very similar to those arising in the field theoretical 
discussion of the quantized scalar and tensor modes of the geometry.
In section \ref{sec3} the quantum treatment of the scalar and tensor modes of the geometry is specifically discussed in a 
unified perspective and by emphasizing those aspects which are germane to the present analysis. In section \ref{sec4} 
the degree of first order coherence is computed and analyzed with particular attention to wavelengths 
larger than the Hubble radius. The intensity correlations are  studied in section \ref{sec5}. 
In sections \ref{sec6} and \ref{sec7} the degree of second-order coherence is computed in different situations 
and always in the framework of the  $\Lambda$CDM scenario. 
The possibilities of a direct estimate of the degree of second-order coherence are also outlined.
 Section \ref{sec8} contains the concluding remarks and the perspectives of forthcoming analyses. 
\renewcommand{\theequation}{2.\arabic{equation}}
\setcounter{equation}{0}
\section{Single mode of the field}
\label{sec2}
By defining the quantum averages with respect to the state $|s\rangle$, the normalized degree of second-order 
coherence\footnote{ For sake of conciseness, the arguments of $\overline{g}^{(2)}$ shall be omitted and it will be understood that 
$\overline{g}^{(2)}$ refers, in this section,  to a single mode of the field and in the zero time-delay limit.}  of Eq. (\ref{EQ4}) can be written as
\begin{equation}
\overline{g}^{(2)} = \frac{\langle s| \hat{a}^{\dagger} \, \hat{a}^{\dagger} \, \hat{a} \hat{a} |s\rangle}{\langle s| \hat{a}^{\dagger} \hat{a} |s \rangle}.
\label{sing1}
\end{equation}
By coarse graining over technical details which will be the subject of the forthcoming sections, it is fair to say that the quantum state of relic phonons (or relic gravitons) belongs to the same class of generalized coherent states which arise in the quantum theory of parametric amplification of Glauber and Mollow (see, e.g. the Hamiltonian of Eq. (3.3) in the first paper of  \cite{mollow}). 
The state $| s \rangle$ introduced in Eq. (\ref{sing1}) will then be the result of the action of a given unitary operator ${\mathcal U}$ on a given initial state $|\mathrm{in} \rangle$, i.e. $| s \rangle =  {\mathcal U} |\mathrm{in} \rangle$. 
In this analogy the unitary operator ${\mathcal U}$ is to be understood as a time evolution operator. 
The averages over $|s\rangle$ can then be made explicit. Since ${\mathcal U}^{-1} = {\mathcal U}^{\dagger}$  and $[\hat{b}, \hat{b}^{\dagger}] =1$,  the linear relation between the creation and annihilation operators ($\hat{a}$, $\hat{a}^{\dagger}$) and 
$(\hat{b}, \hat{b}^{\dagger})$ can be parametrized by the two complex coefficients $c_{\pm}$:
\begin{equation}
\hat{a}= {\mathcal U}^{\dagger} \,\hat{b} \,{\mathcal U} =  c_{+} \,\hat{b} + c_{-}^{*}\,\hat{b}^{\dagger};
\label{sing3}
\end{equation}
since $|c_{+}|^2 - |c_{-}|^2 =1$,  $c_{\pm}$ depend upon three three real numbers (i.e. one amplitude and two phases);
this occurrence is related to an underlying $SU(1,1)$ dynamical symmetry (see e. g. \cite{perelomov} which will be made explicit in section \ref{sec3}.

If the total number of efolds greatly exceeds the maximal number 
of efolds presently accessible by large-scale observations\footnote{In the standard terminology it is customary 
to introduce also $N_{\mathrm{min}}$, i.e. the minimal number of efolds necessary 
to fix the problems of the standard big-bang cosmology} (i.e. $N_{\mathrm{tot}} \gg N_{\mathrm{max}}$), the state 
$|\mathrm{in}\rangle$ coincides, in practice, with the vacuum, i.e. $\hat{b} |\mathrm{in}\rangle =0$. 
In this case the state $|s\rangle$ preserves the minimum uncertainty relations \cite{mollow,perelomov,stoler,yuen} 
(see also \cite{hollenhorst}). 
These states are often dubbed squeezed \cite{yuen,hollenhorst,revsq1,revsq2}
and lead to a specific degree of second-order coherence which will be extremely relevant for the forthcoming considerations. 
The value of $N_{\mathrm{max}}$ 
can be computed once the post-inflationary thermal history is sufficiently well specified (see, e.g. \cite{wein,mgprimer,lid})
 and will be discussed in greater 
detail later on; for the moment it suffices to posit that $N_{\mathrm{max}} \simeq 63$. 
If $N_{\mathrm{tot}} \simeq N_{\mathrm{max}} \simeq N_{\mathrm{min}}$ the statistical properties of $|\mathrm{in}\rangle$ 
can have an impact on the large-scale power spectra as argued in \cite{mg2} and, in this case, $|\mathrm{in} \rangle$ 
does not necessarily coincide with $|0\rangle$. 
Using Eq. (\ref{sing3}) the numerator and denominator of Eq. (\ref{sing1}) are separately given by\footnote{Units $\hbar= c = 1$ will be used throughout.}:
\begin{eqnarray}
\langle s| \hat{a}^{\dagger} \, \hat{a}^{\dagger} \, \hat{a} \hat{a} |s\rangle &=& |c_{-}|^4  \langle \mathrm{in}| \hat{b} \, \hat{b} \, \hat{b}^{\dagger} \hat{b}^{\dagger} |\mathrm{in}\rangle + |c_{+}|^4  \langle \mathrm{in}| \hat{b}^{\dagger} \, \hat{b}^{\dagger} \, \hat{b} \,\hat{b}|\mathrm{in}\rangle 
\nonumber\\
&+& |c_{-}|^2 |c_{+}|^2 \biggl[ \langle \mathrm{in}| \hat{b} \, \hat{b}^{\dagger} \, \hat{b} \hat{b}^{\dagger} |\mathrm{in}\rangle 
\langle \mathrm{in}| \hat{b} \, \hat{b}^{\dagger} \, \hat{b}^{\dagger} \hat{b} |\mathrm{in}\rangle
\nonumber\\
&+& \langle \mathrm{in}| \hat{b}^{\dagger} \, \hat{b} \, \hat{b} \, \hat{b}^{\dagger} |\mathrm{in}\rangle
+ \langle \mathrm{in}| \hat{b}^{\dagger}  \, \hat{b} \, \hat{b}^{\dagger}\, \hat{b}|\mathrm{in}\rangle
\biggr],
\label{sing5}\\
\langle s| \hat{a}^{\dagger} \, \hat{a} |s\rangle^2 &=& \biggl[ |c_{-}|^2 
\langle \mathrm{in}| \hat{b} \, \hat{b}^{\dagger} |\mathrm{in}\rangle + |c_{+}|^2 
\langle \mathrm{in}| \hat{b}^{\dagger} \, \hat{b} |\mathrm{in}\rangle\biggr]^2.
\label{sing6}
\end{eqnarray}
Excluding, for the moment,   the logical possibility that the initial fluctuations have nothing to do with quantum mechanics (see section \ref{sec6}), the state $|\mathrm{in}\rangle$ can be either pure or mixed.

Let us now pause for a moment and let us elaborate on the distinction is between pure (or mixed) states and correlated states which will be relevant for the forthcoming considerations. The purity of a state does not determine, by itself, the degree of second-order coherence. A correlated state (i.e. a state exhibiting a degree of second-order coherence potentially larger than 
a coherent state) can be however modeled in terms of a mixed state whose statistical weights are appropriately chosen. 
Consider, for instance, the following parametrization of the density matrix  
\begin{equation}
\hat{\rho} = |\mathrm{in} \rangle \langle \mathrm{in}|=\sum_{n} P_{n} 
| n \rangle \langle n |, \qquad \sum_{n=0}^{\infty} P_{n} =1.
\label{sing7}
\end{equation}
Thanks to the parametrization of Eq. (\ref{sing7}), the second-order correlation effects of the state 
$|\mathrm{in} \rangle$ will be reflected in the correlation properties of the statistical ensemble defined by the weights 
$P_{n}$  \cite{fano}. A non vanishing initial degree of second-order coherence
\begin{equation}
\overline{g}^{(2)}_{\mathrm{in}} = \frac{\langle \mathrm{in} | \hat{b}^{\dagger} \, \hat{b}^{\dagger} \, \hat{b} \hat{b} |\mathrm{in}\rangle}{\langle \mathrm{in}| \hat{b}^{\dagger} \hat{b} |\mathrm{in} \rangle} \neq 1,
\label{inter2}
\end{equation}
 implies, in terms of the parametrization of Eq. (\ref{sing7}), that the statistical weights must satisfy the condition 
\begin{equation}
\sum_{n=0}^{\infty} n (n -1) P_{n} \neq \biggl[\sum_{n=0}^{\infty} n P_{n}\biggr]^2;
\label{inter3}
\end{equation}
while there are various ways of satisfying the condition (\ref{inter3}) it can be shown that if we want all the 
cumulant moments of the distribution $P_{n}$ to depend only upon the lowest two \cite{karlin}, then  $P_{n}$ must obey the following 
recurrence relation  
\begin{equation}
(n + 1) P_{n + 1}  = ( a + b n) P_{n}, \qquad b \neq 0.
\label{sing7a}
\end{equation}
If $b=0$, then $\overline{g}^{(2)}_{\mathrm{in}}=1$ and $P_{n}$ is given by the standard form of the Poisson distribution with average  multiplicity $\overline{n} = a$. If, however, $b \neq 0$ the generating function of the distribution is simply\footnote{The probability generating function ${\mathcal M}(s)$ can be 
directly obtained from Eq. (\ref{sing7a}) even without knowing the explicit form of $P_{n}$; it suffices 
to multiply the right and left hand sides of Eq. (\ref{sing7a}) by $s^{n}$ and to sum over $n$ 
both sides of the resulting equation. In this way the finite difference equation (\ref{sing7a}) will be 
transformed in a differential equation in $s$ which can be solved by imposing the boundary condition  ${\mathcal M}(1)=1$.}
\begin{equation}
{\mathcal M}(s) = \sum_{n=0}^{\infty} s^{n} P_{n} = \frac{( 1 - b)^{a/b}}{(1 - b s)^{a/b}},
\label{sing8a}
\end{equation}
where clearly ${\mathcal M}(1)=1$ as it must be to be compatible with Eq. (\ref{sing7}).
From Eq. (\ref{sing8a}) the various moments of the distribution are obtained by taking the 
derivatives of Eq. (\ref{sing8a}) at $s=1$. It is useful to parametrize the variance in terms of the ratio between $a$ and $b$ (i.e. 
$\zeta = a/b$) and in terms of $\overline{n}$ (the average multiplicity). 
In terms of these quantities the generating function of Eq. (\ref{sing8a}) can be written as 
\begin{equation}
{\mathcal M}(s) =  \frac{\zeta^{\zeta}}{[\overline{n}( 1 - s) + \zeta]^{\zeta}}, \qquad \frac{D^2}{\overline{n}^2} = \frac{1}{\overline{n}} + \frac{1}{\zeta}.
\label{sing10a}
\end{equation}
where $D^2$ and $\overline{n}$ are defined, respectively, as  $D^2 = {\mathcal M}''(1) + {\mathcal M}'(1) - [{\mathcal M}'(1)]^2$ and $\overline{n} = {\mathcal M}'(1)$; the prime in the two preceding expressions denotes a derivation 
with respect to $s$.  While 
$\overline{n}$ simply denotes the number of particles of the initial state, $D^2$ (and hence $\zeta$) measures the degree 
of second-order coherence of the initial state.
Equation (\ref{sing10a}) shows that the higher moments of the distribution are all expressible, as anticipated, solely in terms of $\zeta$ and $\overline{n}$ since they are given, by definition, as derivatives of  ${\mathcal M}(s)$. The final expression for Eq. (\ref{sing1}) is then given by 
\begin{eqnarray}
\overline{g}^{(2)} &=& \frac{1}{\zeta} \frac{\overline{n}^2 \overline{N}^2 + \overline{n}^2 (\overline{N}+1)^2 + 4 
\overline{N} \overline{n}^2 (\overline{N}+1)}{[ 2 \overline{N} \overline{n} + \overline{n} + \overline{N}]^2}
\nonumber\\
&+& \frac{\overline{N}(\overline{N}+1)[ 4 \overline{n}^2 + 8 \overline{n} +1] + \overline{N}^2 (\overline{n}^2 + 4 \overline{n} + 2 ) + (\overline{N} +1)^2 \overline{n}^2}{[ 2 \overline{N} \overline{n} + \overline{n} + \overline{N}]^2},
\label{sing10}
\end{eqnarray}
 where $\overline{N}= |c_{-}|^2$ and $|c_{+}|^2 = 1 + \overline{N}$.  There are various notable limits of Eq. (\ref{sing10}):
\begin{itemize}
\item{} if $\overline{n} =0$ Eq. (\ref{sing10}) reduces to 
\begin{equation}
\overline{g}^{(2)} = 3 + \frac{1}{\overline{N}},
\label{sing11}
\end{equation}
which is the result expected in the case of the squeezed vacuum state;
\item{} if $\overline{N}= 0$ and $\zeta= 1$, $\overline{g}^{(2)} = 2$ which is the case 
of a thermal state (see, e. g. \cite{sudarshan,loudon});
\item{} if $\overline{N} =0$ and $\zeta \to \infty$ then $\overline{g}^{(2)} = 1$, which is 
the case of a coherent state  (see, e.g. \cite{sudarshan,loudon});
\item{} if $\overline{N} \gg 1$ and $\overline{n} \gg 1$ then $\overline{g}^{(2)} \to 3 ( 1 + 1/\zeta)/2$; thus,  if  $\zeta \to 1$ we shall have that also  $\overline{g}^{(2)} \to 3$ 
as long as $\overline{N} \gg 1$ and $\overline{n} \gg 1$: this result agrees with the well known result concerning the (single mode) squeezed thermal states \cite{knight}.
\end{itemize}
The most general situation corresponds to the case where not only 
$\overline{n} \neq 0$ and $\overline{N} \neq 0$ but also when $\zeta \neq 0$. 
It is appropriate to conclude this discussion with a simple remark on the normal ordering 
in the definition of the degree of second-order coherence. Suppose that we define 
the degree of second-order coherence without resorting to normal ordering; for instance 
we can antinormal order (i.e. $\langle s| \hat{a}\, \hat{a} \, \hat{a}^{\dagger}\, \hat{a}^{\dagger} |s \rangle$) or even 
adopt a mixed kind of ordering (e.g. $\langle s| \hat{a}\, \hat{a}^{\dagger} \, \hat{a}^{\dagger}\, \hat{a} |s \rangle$).
Denoting with $g^{(2)}_{\mathrm{g}}$ the degree of second-order coherence with generic ordering, it can be 
shown that 
\begin{equation}
g^{(2)}_{\mathrm{g}} = \overline{g}^{(2)} + {\mathcal O}(1/\langle N\rangle),
\label{limit}
\end{equation}
where $\langle N \rangle$ schematically denotes the mean number of quanta which depends, ultimately, upon the initial 
state.  

\renewcommand{\theequation}{3.\arabic{equation}}
\setcounter{equation}{0}
\section{Relic phonons and relic gravitons}
\label{sec3}
The simplest setup compatible with the $\Lambda$CDM 
paradigm stipulates that the background geometry is conformally flat with metric tensor 
$\overline{g}_{\mu\nu} = a^2(\tau)\eta_{\mu\nu}$, where $\tau$ denotes the conformal time 
coordinate\footnote{The conventions adopted in the redshifts are such that the present value of the scale 
factor $a_{0}$ is normalized to $1$; note that in section \ref{sec1} $\tau$ denoted, consistently, the time coordinate in Minkowski space }. The scalar and the tensor fluctuations of the geometry 
can be described as \cite{bard1,bard2,bard3} 
\begin{eqnarray}
&& \delta_{(\mathrm{s})} g_{00} = 2 a^2 \phi, \qquad   
 \delta_{(\mathrm{s})} g_{ij} = 2 a^2 \psi \, \delta_{ij},
 \nonumber\\
 && \delta_{(\mathrm{t})} g_{ij} = - a^2 h_{ij}, \qquad \partial_{i}h^{i}_{j} = h_{i}^{i} =0,
 \label{EQQ1}
 \end{eqnarray}
where $\delta_{(\mathrm{s})}$ and $\delta_{(\mathrm{t})}$ denote, respectively, the scalar and the 
the tensor fluctuations of the corresponding quantity; the gauge freedom has been 
completely fixed in Eq. (\ref{EQQ1}) by selecting the conformally Newtonian gauge. 
If the inflationary stage of expansion is driven by a single background scalar field $\varphi$, 
defining with $\delta_{(\mathrm{s})} \varphi$ the scalar fluctuation 
of the inflaton and with $\varphi$ the actions describing the evolution of the scalar 
and tensor modes of the geometry can be written, respectively, as (see, for instance, \cite{mg1,mg1a})
\begin{eqnarray}
&& S_{(\mathrm{s})} = \frac{1}{2} \int \, d^{4} x\, \sqrt{-\overline{g}}
\frac{z^2}{a^2} \overline{g}^{\alpha\beta} \partial_{\alpha} {\mathcal R} \,\partial_{\beta} {\mathcal R},
\label{EQQ2}\\
&&  S_{(\mathrm{t})} = \frac{1}{8 \ell_{\mathrm{P}}^2}   \int \, d^{4} x\, \sqrt{-\overline{g}} \overline{g}^{\alpha\beta} \partial_{\alpha} h_{ij} \,\partial_{\beta} h_{ij},
\label{EQQ3}
\end{eqnarray}
where ${\mathcal R}$ represents the curvature perturbation on comoving 
orthogonal hypersurfaces whose explicit expression, in the case at hand, is
\begin{equation}
{\mathcal R} = - \psi -  \frac{{\mathcal H} \delta_{\mathrm{s}} \varphi}{\partial_{\tau}\varphi}, \qquad z = \frac{a \partial_{\tau} \varphi}{\partial_{\tau} \ln{a}},
\label{EQQ4}
\end{equation}
and ${\mathcal H} = \partial_{\tau} \ln{a}$. Using that $\overline{g}_{\mu\nu} = a^2(\tau) \eta_{\mu\nu}$, Eqs. (\ref{EQQ2}) and 
(\ref{EQQ3}) can be written as 
\begin{eqnarray}
&& S_{(\mathrm{s})} = \frac{1}{2} \int d^{4} x \,\,\eta^{\alpha\beta} 
\partial_{\alpha} {\mathcal R}\, \partial_{\beta} {\mathcal R} \, z^2,
\label{EQQ5}\\
&& S_{(\mathrm{t})} = \frac{1}{2} \int d^{4} x \,\,\eta^{\alpha\beta} 
\partial_{\alpha} h \,\partial_{\beta} h \, a^2.
\label{EQQ6}
\end{eqnarray}
Equation (\ref{EQQ6}) holds for each of the two tensor 
polarizations (see, e. g. \cite{mg1,fordp}) having defined $h_{\xi} = \sqrt{2} \ell_{\mathrm{P}} h$ with $\xi= \oplus,\,\otimes$ and $\ell_{\mathrm{P}} = 1/\sqrt{8 \pi G}$ (see also Eq. (\ref{DS4}) where the reduced Planck mass $\overline{M}_{\mathrm{P}} = \ell_{\mathrm{P}}^{-1}$ enters the definition of the slow-roll parameters).  Indeed $h_{ij}(\vec{x},\tau)$ can be decomposed as 
\begin{equation}
h_{ij}(\vec{x},\tau) = \sum_{\xi} q^{(\xi)}_{ij} h_{\lambda}(\vec{x},\tau),
\label{EQQ7}
\end{equation}
where $q^{\oplus}_{ij} 
= (\hat{a}_{i} \hat{a}_{j} - \hat{b}_{i} \hat{b}_{j})$ and 
$q^{\otimes}_{ij}= (\hat{a}_{i} \hat{b}_{j} + \hat{a}_{j} \hat{b}_{i})$; defining with $\hat{k}$ the direction of propagation 
of the wave, $\hat{a}$, $\hat{b}$ and $\hat{k}$ form a triplet of mutually orthogonal unit vectors. Probably 
the first paper mentioning quantum mechanics as a possible source of large-scale inhomogeneities, though 
not in the framework of any inflationary hypothesis, is the one of Sakharov \cite{sakharov}.
The emphasis on the action for the normal mode of the scalar 
fluctuations (i.e. scalar phonons) appeared in a paper by Lukash \cite{luk} in the context 
of fluid models. Later on different authors applied it to scalar field matter with particular attention to the 
quantization of the fluctuations \cite{KS,chibisov} (see also \cite{strokov}).  The form of the actions is the one derived in \cite{mg1}. The scalar  and tensor fluctuations of the geometry can be canonically quantized; after introducing the appropriate normal modes 
\begin{equation}
\mu(\vec{x},\tau) = a(\tau) h(\vec{x},\tau), \qquad \nu(\vec{x},\tau) = z(\tau) {\mathcal R}(\vec{x},\tau),
\label{EQQ8}
\end{equation}
the tensor and scalar Lagrangian densities become, respectively,
\begin{eqnarray}
&& {\mathcal L}_{(\mathrm{t})}(\vec{x},\tau) = \frac{1}{2} \biggl[(\partial_{\tau} \mu)^2 + (\partial_{\tau} \ln{a})^2 \mu^2 
- 2 (\partial_{\tau} \ln{a}) \, \mu \partial_{\tau} \mu - (\partial_{i} \mu)^2\biggr],
\label{EQQ9}\\
&& {\mathcal L}_{(\mathrm{s})}(\vec{x},\tau) = \frac{1}{2} \biggl[(\partial_{\tau} \nu)^2 + (\partial_{\tau} \ln{z})^2 \mu^2 
- 2 (\partial_{\tau} \ln{z}) \, \nu \partial_{\tau} \nu - (\partial_{i} \nu)^2\biggr],
\label{EQQ10}
\end{eqnarray}
whose associated canonical momenta 
\begin{equation}
\pi_{(\mathrm{t})}= \partial_{\tau}\mu- (\partial_{\tau} \ln{a}) \mu, \qquad 
\pi_{(\mathrm{s})}= \partial_{\tau}\nu - (\partial_{\tau} \ln{z}) \nu,
\label{EQQ13}
\end{equation}
can be used to derive the canonical Hamiltonians 
\begin{eqnarray}
&& H_{(\mathrm{t})} = \frac{1}{2} \int d^{3} x \biggl[ \pi_{(\mathrm{t})}^2 + 2 (\partial_{\tau} \ln{a}) \pi_{(\mathrm{t})} \mu + (\partial_{i} \mu)^2 \biggr],
\label{EQQ11}\\
&& H_{(\mathrm{s})} = \frac{1}{2} \int d^{3} x \biggl[ \pi_{(\mathrm{s})}^2 + 2 (\partial_{\tau} \ln{z}) \pi_{(\mathrm{s})} \nu + (\partial_{i} \nu)^2 \biggr].
\label{EQQ12}
\end{eqnarray}
Since Eqs. (\ref{EQQ11}) and (\ref{EQQ12}) have the same canonical 
structure, the two problems can be treated simultaneously by resorting to the Hamiltonian:
\begin{equation}
\hat{H}(\tau) = \frac{1}{2} \int d^{3} x \biggl[ \hat{\pi}^2 - 2 i \lambda ( \hat{\pi} \hat{\Phi} + 
\hat{\Phi} \hat{\pi}) + \partial_{k} \hat{\Phi} \partial^{k} \hat{\Phi}\biggr],
\label{EQQ14}
\end{equation}
where $\hat{\Phi}$ and $\hat{\pi}$ are the two canonically conjugate field operators. In Eq. (\ref{EQQ14}) 
$\lambda= i (\partial_{\tau} \ln{a})/2$ (in the case of the tensor modes) 
and $\lambda=  i (\partial_{\tau} \ln{z})/2$ (in the case of the scalar modes). Similarly 
$\hat{\pi}$ will coincide either with $\hat{\pi}_{(\mathrm{t})}$ (in the case of the tensor modes) or with 
$\hat{\pi}_{\mathrm{(s)}}$ (in the case of the scalar modes). The Fourier representation of the 
field operators can be written as 
\begin{equation}
\hat{\Phi}(\vec{x},\tau) = \frac{1}{\sqrt{V}} \sum_{\vec{p}} \hat{\Phi}_{\vec{p}}(\tau)\, e^{- i \vec{p}\cdot \vec{x}},\qquad 
\hat{\pi}(\vec{x},\tau) = \frac{1}{\sqrt{V}} \sum_{\vec{p}} \hat{\pi}_{\vec{p}}(\tau)\, e^{- i \vec{p}\cdot \vec{x}},
\label{EQQ14a}
\end{equation}
where $V$ represents a fiducial (normalization) volume.
In the continuum limit  we will have $\sum_{\vec{k}} \to V \int d^{3} k/(2\pi)^3$ and 
the canonical commutation relations impose, in Fourier space, 
$[ \hat{\Phi}_{\vec{k}}, \hat{\pi}_{\vec{p}}^{\dagger}] = i \delta^{(3)}(\vec{k} - \vec{p})$
where $\hat{\Phi}_{\vec{k}}^{\dagger} = \hat{\Phi}_{-\vec{k}}$ and 
$\hat{\pi}_{\vec{k}}^{\dagger} = \hat{\pi}_{-\vec{k}}$ because of the hermiticity 
of the corresponding field operators in real space. Introducing creation 
and annihilation operators obeying $[\hat{a}_{\vec{k}}, \hat{a}_{\vec{p}}^{\dagger}] = 
\delta^{(3)}(\vec{k} - \vec{p})$,  the field operators and the canonical momenta can be expressed as
\begin{equation}
\hat{\Phi}_{\vec{p}} = \frac{1}{\sqrt{2 p}} ( \hat{a}_{\vec{p}} + \hat{a}_{-\vec{p}}^{\dagger}),\qquad \hat{\pi}_{\vec{p}} = - i \sqrt{\frac{p}{2}} ( \hat{a}_{\vec{p}} - \hat{a}_{-\vec{p}}^{\dagger}).
\label{EQQ15}
\end{equation}
Inserting Eq. (\ref{EQQ15}) into Eq. (\ref{EQQ14}) the resulting Hamiltonian in the continuum limit is 
\begin{equation}
\hat{H}(\tau) =2  \int d^{3} p \biggl\{\,\,p \,\, {\mathcal K}_{0}(\vec{p})  + \biggl[ \lambda^{*}(\tau) {\mathcal K}_{-}(\vec{p}) + \lambda(\tau) {\mathcal K}_{+}(\vec{p})\biggr]\biggr\},
\label{EQQ16}
\end{equation}
where the operators ${\mathcal K}_{\pm}(\vec{p})$ and ${\mathcal K}_{0}(\vec{p})$ 
\begin{equation}
{\mathcal K}_{+}(\vec{p}) = \hat{a}_{\vec{p}}^{\dagger} \,\hat{a}_{-\vec{p}}^{\dagger},\qquad {\mathcal K}_{-}(\vec{p}) = \hat{a}_{\vec{p}}\, \hat{a}_{-\vec{p}},
\qquad {\mathcal K}_{0}(\vec{p}) = \frac{1}{2}\biggl[ \hat{a}_{\vec{p}}^{\dagger}\, \hat{a}_{\vec{p}} + \hat{a}_{-\vec{p}} \,\hat{a}_{-\vec{p}}^{\dagger}\biggr],
\label{EQQ17}
\end{equation}
 satisfy the commutation relations of the $SU(1,1)$ Lie algebra, i.e. 
\begin{equation}
 [{\mathcal K}_{-}(\vec{p}) , {\mathcal K}_{+}(\vec{q})] = 2 \,{\mathcal K}_{0}(\vec{p}) \,
\delta^{(3)}(\vec{p} - \vec{q}),\qquad [{\mathcal K}_{0}(\vec{p}), {\mathcal K}_{\pm}(\vec{q})] = \pm \,{\mathcal K}_{\pm}(\vec{p})\, \delta^{(3)}(\vec{p}-\vec{q}).
\label{EQQ18}
\end{equation}
The group $SU(1,1)$ is not a symmetry group of the Hamiltonian 
of the problem but the $SU(1,1)$ algebra can be viewed, in the terminology of 
\cite{solomon} (see also \cite{mg4}), as the spectrum generating algebra insofar as  
the total (generalized) charge does commute with all the generators of the group (as well as with the total 
Hamiltonian) while the total number of particles does commute with the charge but not 
with the full Hamiltonian.  Owing to the group structure (\ref{EQQ18}) and to the specific form of the 
Hamiltonian of Eq. (\ref{EQQ16}), the multiparticle final state can be obtained by applying 
to the initial state  $|\Psi_{i}(\vec{p})\rangle$ the product of two unitary operators $\Xi(\varphi_{p})$ and $\Sigma(\sigma_{p})$:
\begin{equation}
|\Psi_{f}(\vec{p})\rangle = \Xi(\varphi_{p}) \, \Sigma(\sigma_{p}) |\Psi_{i}(\vec{p})\rangle, \qquad |\Psi_{f}\rangle = \prod_{\vec{p}} |\Psi_{f}(\vec{p})\rangle,
\label{EQQ19}
\end{equation}
where the unitary operators are defined as\footnote{Note that $\varphi_{p}$ should not be confused 
with $\varphi$ (denoting the inflaton field). This confusion cannot actually arise 
since it is clear that $\varphi_{p}$ is a momentum-dependent phase.}
\begin{equation}
{\mathcal R}(\varphi_{p}) = \exp{[ - 2 \,i \,\varphi_{p}\, K_{0}(\vec{p})]}, \qquad \Sigma(\sigma_{p}) = \exp{[\sigma_{p}^{*}\, K_{-}(\vec{p}) - 
\sigma_{p}\, K_{+}(\vec{p})]},
\label{EQQ20}
\end{equation}
with $\sigma_{p} = r_{p} e^{i \gamma_{p}}$ and $\alpha_{p} = (2\varphi_{p} -\gamma_{p})$; the time evolution of the 
variables $r_{p}(\tau)$, $\varphi_{p}(\tau)$  and $\alpha_{p}(\tau)$ is given by 
\begin{eqnarray}
&& \frac{d r_{p}}{d\tau} = 2 i \lambda \cos{\alpha_{p}}, 
\nonumber\\
&& \frac{d\varphi_{p}}{d\tau} = p  - 2 i \lambda \tanh{r_{p}} \sin{\alpha_{p}},
\nonumber\\
&& \frac{d \alpha_{p}}{d\tau}= 2 p - 4 i \lambda \frac{\sin{\alpha_{p}}}{\tanh{2 r_{p}}}.
\label{EQQ21}
\end{eqnarray}
Equations (\ref{EQQ21}) are symmetric for $r_{p} \to - r_{p}$ and $\lambda \to - \lambda$. 
The corresponding Hamiltonian of Eqs. (\ref{EQQ14}) is actually symmetric for 
\begin{equation}
z\to \frac{1}{z}\qquad \hat{\pi}_{\vec{k}} \to - k \hat{\Phi}_{\vec{k}}, \qquad \hat{\Phi}_{\vec{k}} \to \frac{1}{k} \hat{\pi}_{\vec{k}},
\label{EQQ21a}
\end{equation}
and analogously in the tensor case for $a\to 1/a$. The transformation of Eq. (\ref{EQQ21a}) 
is related to electric-magnetic duality 
\cite{DT,deser,mg5} in conformally flat background geometries when the $a$ and $z$ are replaced 
by the (dynamical) gauge coupling.
The relation between the Schr\"odinger and the Heisenberg descriptions is 
easily worked out by appreciating that 
\begin{eqnarray}
&& \hat{\Phi}_{\vec{k}}(\tau) = f_{k}(\tau) \hat{a}_{\vec{k}}(\tau_{0}) + f_{k}^{\ast} \hat{a}^{\dagger}_{- \vec{k}}(\tau_{0}),
\label{H1}\\
&& \hat{\pi}_{\vec{k}}(\tau) = g_{k}(\tau) \hat{a}_{\vec{k}}(\tau_{0}) + g_{k}^{\ast} \hat{a}^{\dagger}_{- \vec{k}}(\tau_{0}).
\label{H2}
\end{eqnarray}
In section \ref{sec2} the operators at the initial time $\tau_{0}$ have been denoted with $\hat{b}$; thus, accounting 
for the momentum dependence, we will denote $\hat{b}_{\vec{k}} = \hat{a}_{\vec{k}}(\tau_{0})$, 
$b_{-\vec{k}}^{\dagger} = \hat{a}^{\dagger}_{- \vec{k}}(\tau_{0})$. 
The evolution of the mode functions $f_{k}(\tau)$ and $g_{k}(\tau)$ 
can be obtained from the evolution equations in the Heisenberg 
description:
\begin{equation}
i \partial_{\tau} \hat{\Phi}= [\hat{\Phi}, \hat{H}], \qquad i \partial_{\tau}\hat{\pi} = [ \hat{\pi}, \hat{H}], 
\label{H2a}
\end{equation}
and they are given by 
\begin{equation}
 \partial_{\tau} f_{k}  = g_{k} + \partial_{\tau} \ln{a} f_{k}, \qquad \partial_{\tau} g_{k} = - k^2 f_{k} - (\partial_{\tau}\ln{a})g_{k}
\label{H2b}
\end{equation}
for the tensor case and by 
\begin{equation}
 \partial_{\tau} \tilde{f}_{k}  = \tilde{g}_{k} + \partial_{\tau} \ln{z} \tilde{f}_{k}, \qquad \partial_{\tau} \tilde{g}_{k} = - k^2 \tilde{f}_{k} - (\partial_{\tau}\ln{z}) \tilde{g}_{k}
\label{H2c}
\end{equation}
for the scalar case.  Both sets of mode functions are subjected to the following Wronskian normalization for any $\tau$ 
\begin{eqnarray}
&& \tilde{f}_{p}(\tau) \, \tilde{g}_{p}^{*}(\tau) - \tilde{f}_{p}^{*}(\tau) \, \tilde{g}_{p}(\tau) = i,
\label{WR1}\\
&& f_{p}(\tau) g_{p}^{*}(\tau) - f_{p}^{*}(\tau) g_{p}(\tau) = i.
\label{WR2}
\end{eqnarray}

\renewcommand{\theequation}{4.\arabic{equation}}
\setcounter{equation}{0}
\section{First-order coherence}
\label{sec4}
The degree of first-order coherence is measured by the following normalized 
correlation function:
\begin{equation}
g^{(1)}(\vec{x}, \vec{y}; \tau) = \frac{\langle \hat{\Phi}(\vec{x},\tau) \hat{\Phi}(\vec{y},\tau)\rangle }{
\sqrt{ \langle |\hat{\Phi}(\vec{x},\tau)|^2\rangle\,\langle|\hat{\Phi}(\vec{y},\tau)|^2 \rangle}}.
\label{F1A}
\end{equation}
In the usual quantum optical setup provided by standard interferometers (such as Michelson, Mach-Zehnder or Sagnac) 
the $\hat{\Phi}$-operators are replaced either by electric field operators or even by classical fields. 
In Young interferometry the electric fields are split, time delayed and then 
recombined on a screen. It is easy to show that the analog of Eq. (\ref{F1A}) is nothing but the visibility, i.e. the normalized 
difference between the maximal and the minimal intensity of the light detected on the Young screen.

In quantum theory, the degree of first-order coherence is fully determined 
by the averaged multiplicity of the initial state. Let us suppose that the initial 
state is characterized by an averaged multiplicity $\overline{n}_{q}$ per each field mode. Following 
the notations employed in section \ref{sec2} but accounting for the momentum dependence we will have
\begin{equation}
\langle \mathrm{in} |\hat{b}^{\dagger}_{\vec{q}} \, \hat{b}_{\vec{p}} |\mathrm{in} \rangle = 
\overline{n}_{q} \delta^{(3)}(\vec{q} - \vec{p}).
\label{F2A}
\end{equation}
Given the average occupation number of the initial state, Eq. (\ref{F1A}) can be computed in explicit terms. 
Recalling Eqs. (\ref{EQQ14a}) and (\ref{EQQ15}) the two point function is simply given by
\begin{eqnarray}
&& \langle \hat{\Phi}(\vec{x},\tau) \hat{\Phi}(\vec{y},\tau)\rangle = \frac{1}{4\pi^2} 
\int q\,d q \biggl[ \cosh{2 r_{q}}  - \cos{\alpha_{q}} \sinh{2 r_{q}}\biggr] ( 2 \overline{n}_{q} +1) j_{0}(q r),
\label{F4A}\\
&& \langle :\, \hat{\Phi}(\vec{x},\tau) \hat{\Phi}(\vec{y},\tau)\, : \rangle = \frac{1}{2\pi^2} 
\int q\,d q \biggl[ \cosh{2 r_{q}}  - \cos{\alpha_{q}} \sinh{2 r_{q}}\biggr]  \overline{n}_{q}\, j_{0}(q r),
\label{F5A}
\end{eqnarray}
where the normal ordered case has been included for comparison and where $j_{0}(q r)$ is the zeroth-order spherical 
Bessel function \cite{abr,tric}. To deduce Eqs. (\ref{F4A}) and (\ref{F5A}) it is useful to recall that
\begin{equation} 
\hat{a}_{\vec{q}} = e^{- i \varphi_{q}}\biggl[ \cosh{r_{q}}\, \hat{b}_{\vec{q}} - e^{i \gamma_{q}} \sinh{r_{q}} \hat{b}^{\dagger}_{-\vec{q}}\biggr],
\label{SC6}
\end{equation}
where, as in Eqs. (\ref{EQQ19})-(\ref{EQQ20}), $\gamma_{q} = (2 \varphi_{q} -\alpha_{q})$. 
Since $\sinh^2{r_{q}} = \overline{N}_{q}$ is the averaged multiplicity 
of the squeezed vacuum state, the expression 
appearing inside the square brackets in Eqs. (\ref{F4A}) and 
(\ref{F5A}) can also be written as 
\begin{equation}
\cosh{2 r_{q}}  - \cos{\alpha_{q}} \sinh{2 r_{q}} = 2\overline{N}_{q} + 1 -
2 \sqrt{\overline{N}_{q} (\overline{N}_{q} +1)} \cos{\alpha_{q}}.
\label{F6A}
\end{equation}
The solution of either Eqs. (\ref{EQQ21}) or Eqs. (\ref{H2b})--(\ref{H2c}) leads to 
an even more explicit form of the degree of first-order coherence which will be indirectly mentioned in section \ref{sec7}. 
In spite of the statistical properties of the initial state and in spite of the operator ordering Eqs. (\ref{F4A}) and 
(\ref{F5A}) imply that 
\begin{equation}
\lim_{q r \to 0} g^{(1)}(r,\tau) = 1, \qquad r = |\vec{x} - \vec{y}|, \qquad q = |\vec{q}|.
\label{F3A}
\end{equation}
Concerning the limit of Eq. (\ref{F3A}) few comments are in order. 
Mathematically the correct limit to be implemented is $r\to 0$ since the correlation function is the result of an integral
over the comoving three-momentum; at the same time the physical limit, as indicated in Eq. (\ref{F3A}) 
is $q r \ll 1$ (and also, as we shall see in the case of second-order correlations, $|k \tau|\ll 1$). 
From the explicit expression of $g^{(1)}(r,\tau)$ it is also clear that the integrals over $q$ might not always 
be convergent. Still the result of the limit holds because, for $r \to 0$ the divergent contributions 
in the numerator and in the denominator exactly cancel.

The result of Eq. (\ref{F3A}) can be dubbed by saying that the relic gravitons and the relic phonons are always first-order coherent
\footnote{The terminology 
``first-order coherence" (and later on of ``second-order coherence") is the one borrowed 
from the Glauber theory of optical coherence as formulated in \cite{glauber1,glauber2}. In section \ref{sec1} 
the basics of Glauber approach to optical fields have been reviewed and will now be applied in the present and in the following sections to the case of relic phonons and relic gravitons.} irrespective of the statistical properties of the initial state which could be rather different such as  a mixed or a pure state. In both 
cases the density operator of the initial state can be defined in the most appropriate 
basis, for instance a Fock basis or a coherent state basis. For a thermal (or chaotic) ensemble
the density matrix can be written, in the Fock basis, as 
\begin{equation}
\hat{\rho} = \sum_{\{n\}} P_{\{n \}} |\{n \}\rangle \langle \{n \}|,
\qquad P_{\{n\}} = \prod_{\vec{k}} \frac{\overline{n}_{\vec{k}}^{n_{\vec{k}}}}{( 1 + \overline{n}_{\vec{k}})^{n_{\vec{k}} + 1 }},
\label{F7A}
\end{equation}
where, in analogy with the notations employed in section \ref{sec2}, $\overline{n}_{\vec{k}} = \mathrm{Tr}[ \hat{\rho} \, \hat{a}_{\vec{k}}^{\dagger}\, \hat{a}_{\vec{k}}]$ is the average occupation number of each Fourier mode and, following 
the standard notation, $ |\{n \}\rangle = |n_{\vec{k}_{1}} \rangle \, |  |n_{\vec{k}_{2}} \rangle \, |  |n_{\vec{k}_{3}} \rangle...$ 
where the ellipses stand for all the occupied modes of the field. The density 
matrix always describes a mixed state but the  $\overline{n}_{\vec{k}}$  should not be necessarily identified 
with the Bose-Einstein occupation number. In the case of a multimode coherent state the density matrix can instead be written as  
\begin{eqnarray}
\hat{\rho} &=& |\{\beta\}\rangle \langle \{\beta\}|, \qquad |\{\beta\} \rangle = \prod_{\vec{k}} \,|\{\beta_{\vec{k}}\}\rangle,
\nonumber\\
|\{\beta_{\vec{k}}\}\rangle &=& e^{- |\beta_{\vec{k}}|^2/2} \sum_{n_{\vec{k}}} \frac{\beta_{\vec{k}}^{n_{\vec{k}}}}{\sqrt{n_{\vec{k}} \,!}} \, | n_{\vec{k}} \rangle .
\label{F8A}
\end{eqnarray}
The initial state of Eq. (\ref{F8A}) is pure and its 
averaged multiplicity per Fourier mode is given by $\overline{n}_{k} = |\beta_{k}|^2$. Hence Eq. (\ref{F3A}) holds both for the state of Eq. (\ref{F7A}) and for the state of Eq. (\ref{F8A}) as well as for all the possible initial states (either pure or mixed).
This simply means that the degree of first-order coherence is only sensitive to the particle content of the initial state but not to its statistical properties. Initial states exhibiting a high degree of correlation cannot be distinguished just by looking at the analog of Young interferometry.  A given multiparticle 
state can always be projected on the coherent state basis \cite{sudarshan}.

The idea is to write the multiparticle density matrix of a mixed state in the basis of a multimode coherent state using the 
overcompleteness of $|\{ \gamma \}\rangle$ 
\begin{eqnarray}
\hat{\rho} &=& \int P(\{\gamma\}) \, |\{ \gamma \}\rangle \langle \{ \gamma \} | d \mu \{\gamma\}
\label{F9A}\\
d \mu \{\gamma\} &=& \Pi_{\vec{k}} \biggl[ \frac{1}{\pi} d^{2} \, \gamma_{\vec{k}}\biggr],
\label{F10A}
\end{eqnarray}
where $P(\{\gamma\})$ is the phase-space functional. The representation of Eq. (\ref{F9A}) cannot be more singular than 
a Dirac delta function and it should also be positive semi-definite: these two properties are not satisfied 
by any quantum state. For instance, in the case of the density matrices of Eqs. (\ref{F7A}) and (\ref{F8A}) 
the corresponding phase-space functionals are:
\begin{eqnarray}
P(\{\gamma\}) &=& \prod_{\vec{k}} \, \frac{1}{\overline{n}_{\vec{k}}}\,\, \exp{[- |\gamma_{\vec{k}}|^2/\overline{n}_{\vec{k}}]},
\label{F11A}\\
P(\{\gamma\}) &=& \prod_{\vec{k}} \delta^{(2)}(\gamma_{\vec{k}} -\beta_{\vec{k}}).
\label{F12A}
\end{eqnarray}
There are indeed states leading to a phase-space functional which is more singular than a delta function (the delta function 
case corresponding to a coherent state) such as the squeezed states and various of their generalizations 
\cite{loudon2,stenholm}. 
As it will be clear in what follows, the method of the phase-space functional will just be mentioned 
as a cross-check in the calculation of some expectation values involving the initial state of the relic phonons and of the 
relic gravitons. In the cases where this technique will be employed, the $P$-representation will always be well defined and regular.
When normal ordering is imposed, 
in the coherent state basis defined by Eqs. (\ref{F9A})--(\ref{F10A}) and by Eqs. (\ref{F11A})--(\ref{F12A}),
 the quantum averages are replaced by averages 
over complex numbers weighted by the phase space functional related to the so-called Glauber-Sudarshan 
$P$-representation \cite{sudarshan,loudon,mandel}. The latter properties goes under the name 
of optical equivalence theorem and greatly simplifies the calculations of field correlators provided 
the density matrix of the quantum states involved in the average can be represented in the coherent state 
basis with $P$-distribution not more singular than a Dirac delta function \cite{sudarshan,loudon,mandel}.
This observation can be used to check the results of various correlators in the limit of large occupation numbers 
since, in this limit, the ordering of the higher-order correlators is immaterial as discussed at the end of section \ref{sec2}.

The main result of this section can be summarized by saying that the field describing the relic phonons and the relic gravitons is always first-order coherent when the 
relevant wavelengths are larger than the Hubble radius at each corresponding 
epoch. Furthermore, given the explicit form of Eq. (\ref{F4A}) and (\ref{F5A}),
we can also conclude that $ 0 \leq g^{(1)}(r,\tau) \leq 1$, i.e. the field is 
first-order coherent in the large-scale limit and partially coherent 
for smaller scales. 

\renewcommand{\theequation}{5.\arabic{equation}}
\setcounter{equation}{0}
\section{Intensity correlations}
\label{sec5}
The degree of second-order coherence is defined as in Eq. (\ref{EQ3}), i.e. 
\begin{equation}
g^{(2)}(\vec{x},\vec{y}; \tau) = \frac{\langle \hat{{\mathcal I}}(\vec{x},\tau) \, \hat{{\mathcal I}}(\vec{y},\tau)\rangle}{\langle\, 
\hat{{\mathcal I}}(\vec{x},\tau)\,\rangle \langle\, 
\hat{{\mathcal I}}(\vec{y},\tau)\,\rangle},
\label{SC1}
\end{equation}
where $\hat{{\mathcal I}}(\vec{x},\tau) = \hat{\Phi}^2(\vec{x},\tau)$ and 
$\hat{{\mathcal I}}(\vec{y},\tau) = \hat{\Phi}^2(\vec{y},\tau)$. 
As anticipated in the introductory section Eq. (\ref{SC1}) differs from the quantum degree of second-order 
coherence appearing in Refs. \cite{sudarshan,loudon,mandel}.  In quantum optics the intensities 
are measured by phototubes and the correlation is proportional to the transition rate for a joint absorption of photons 
at the two points. The treatment of the photoelectric effect shows that the transition amplitude is proportional 
to the matrix element of $\hat{E}^{(+)}(\vec{y}, \tau_{2}) \, \hat{E}^{(+)}(\vec{x}, \tau_{1})$; accordingly the 
degree of second-order coherence is defined, in quantum optics, as\footnote{As in section \ref{sec2}, 
$\overline{g}^{(2)}$ denote the normal ordered degree of second-order coherence while in 
Eq. (\ref{SC1}) the bar has been omitted.}\cite{sudarshan,loudon,mandel}
\begin{equation}
\overline{g}^{(2)}(\vec{x},\vec{y}; \tau_{1}, \tau_{2}) = \frac{\langle : \hat{E}^{(-)}(\vec{x}, \tau_{1}) \, \hat{E}^{(-)}(\vec{y}, \tau_{2})
\hat{E}^{(+)}(\vec{y}, \tau_{2}) \, \hat{E}^{(+)}(\vec{x}, \tau_{1}):\rangle}{\langle:  \hat{E}^{(-)}(\vec{x}, \tau_{1})  \hat{E}^{(+)}(\vec{x}, \tau_{1}):\rangle \langle: \hat{E}^{(-)}(\vec{y}, \tau_{2})  \hat{E}^{(+)}(\vec{y}, \tau_{2}):\rangle},
\label{SC2}
\end{equation}
where $\hat{E}^{(-)}(\vec{x}, \tau)$ and  $\hat{E}^{(+)}(\vec{x}, \tau)$ denote, respectively, the 
negative and the positive frequency parts of the electric field operator for a single polarization. By rewriting Eq. 
(\ref{SC2}) in the notation of the present paper the quantum 
degree of second-order coherence becomes, in the case $\tau_{1} = \tau_{2} = \tau$, 
\begin{equation}
\overline{g}^{(2)}(\vec{x},\vec{y}; \tau) = \frac{\langle :\hat{\Phi}^{(-)}(\vec{x}, \tau) \, \hat{\Phi}^{(-)}(\vec{y}, \tau)
\hat{\Phi}^{(+)}(\vec{y}, \tau) \, \hat{\Phi}^{(+)}(\vec{x}, \tau):\rangle}{\langle\,: \hat{\Phi}^{(-)}(\vec{x}, \tau)  
\hat{\Phi}^{(+)}(\vec{x}, \tau):\rangle \langle : \hat{\Phi}^{(-)}(\vec{y}, \tau)  \hat{\Phi}^{(+)}(\vec{y}, \tau):\rangle},
\label{SC3}
\end{equation}
where, as above, $\hat{\Phi}^{(-)}(\vec{x}, \tau)$ and  $\hat{\Phi}^{(+)}(\vec{x}, \tau)$ denote the negative and the positive 
frequency parts of $\hat{\Phi}$.
Equations (\ref{SC1}) and (\ref{SC3})  are technically different but physically equivalent. The numerical
value of the normalized degree of second-order coherence (in the zero time-delay limit) 
can differ between Eqs. (\ref{SC1}) and (\ref{SC3}) for a given quantum state. However, as 
discussed in Eq. (\ref{limit}) these difference vanish either when the number of particles of the initial 
state is large or when the number of produced particles is large.  Barring for specific numerical differences 
which are relevant in the limit of small occupation numbers, Eq. (\ref{SC3}) shall be primarily considered; if appropriate, the relations 
of the obtained results with the normal ordered definition shall be swiftly mentioned. 
The degree of second-order coherence given in Eq. (\ref{SC1}) can be estimated as 
\begin{equation}
\langle \hat{{\mathcal I}}(\vec{x},\tau) \hat{{\mathcal I}}(\vec{y},\tau)\rangle = \frac{1}{V^2} \sum_{\vec{q}} \,\sum_{\vec{p}}\,
 \sum_{\vec{q}^{\prime}} \,\sum_{\vec{p}^{\prime}} {\mathcal F}(q,\, p,\, q^{\prime},\, p^{\prime};\, \vec{x},\, \vec{y},\, \tau)
\label{SC4}
\end{equation}
where the expression ${\mathcal F}(q,\, p,\, q^{\prime},\, p^{\prime};\, \vec{x},\, \vec{y},\, \tau)$ is given by 
\begin{eqnarray}
&& \biggl\{ \langle \hat{a}_{\vec{q}} \, \hat{a}_{\vec{p}}\, 
\hat{a}^{\dagger}_{\vec{q}\,'}\, \hat{a}^{\dagger}_{\vec{p}\,'}\,\rangle 
e^{ - i (\vec{q} + \vec{p}) \cdot \vec{x} + i (\vec{q}\,' + \vec{p}\,')\cdot \vec{y}} + \langle \hat{a}^{\dagger}_{\vec{q}} \, \hat{a}^{\dagger}_{\vec{p}}\, 
\hat{a}_{\vec{q}\,'}\, \hat{a}_{\vec{p}\,'}\,\rangle 
e^{ i (\vec{q} + \vec{p}) \cdot \vec{x} - i (\vec{q}\,' + \vec{p}\,')\cdot \vec{y}}
\nonumber\\
&&  \langle \hat{a}_{\vec{q}} \, \hat{a}^{\dagger}_{\vec{p}}\, 
\hat{a}_{\vec{q}\,'}\, \hat{a}_{\vec{p}\,'}^{\dagger}\,\rangle
e^{- i (\vec{q} - \vec{p}) \cdot \vec{x} - i (\vec{q}\,' - \vec{p}\,')\cdot \vec{y}}
+ \langle \hat{a}_{\vec{q}} \, \hat{a}^{\dagger}_{\vec{p}}\, 
\hat{a}_{\vec{q}\,'}^{\dagger}\, \hat{a}_{\vec{p}\,'}\,\rangle 
e^{- i (\vec{q} - \vec{p}) \cdot \vec{x} + i (\vec{q}\,' - \vec{p}\,')\cdot \vec{y}}
\nonumber\\
&& \langle \hat{a}_{\vec{q}}^{\dagger} \, \hat{a}_{\vec{p}}\, 
\hat{a}_{\vec{q}\,'}\, \hat{a}_{\vec{p}\,'}^{\dagger}\,\rangle
e^{i (\vec{q} - \vec{p}) \cdot \vec{x} - i (\vec{q}\,' - \vec{p}\,')\cdot \vec{y}}
+  \langle \hat{a}_{\vec{q}}^{\dagger} \, \hat{a}_{\vec{p}}\, 
\hat{a}_{\vec{q}\,'}^{\dagger}\, \hat{a}_{\vec{p}\,'}^{\dagger}\,\rangle
e^{i (\vec{q} - \vec{p}) \cdot \vec{x} + i (\vec{q}\,' - \vec{p}\,')\cdot \vec{y}}\biggr\}.
\label{SC5}
\end{eqnarray}
Inserting Eq. (\ref{SC6}) into Eq. (\ref{SC5}) various averages will appear such as 
\begin{equation} 
\langle \hat{b}^{\dagger}_{\vec{q}} \, \hat{b}^{\dagger}_{\vec{p}}\, \hat{b}_{\vec{q}^{\,\prime}} \, \hat{b}_{\vec{p}^{\,\prime}} \rangle, 
\qquad \langle \hat{b}_{\vec{q}} \, \hat{b}^{\dagger}_{\vec{p}}\, \hat{b}^{\dagger}_{\vec{q}^{\,\prime}} \, \hat{b}_{\vec{p}^{\,\prime}} \rangle,\,...
\label{SC7}
\end{equation}
where  the ellipses stand for the four remaining permutations. Each o the averages Eq. (\ref{SC7}) are evaluated 
using the density matrix of the initial state, i.e. for instance
\begin{equation}
\langle \hat{b}^{\dagger}_{\vec{q}} \, \hat{b}^{\dagger}_{\vec{p}}\, \hat{b}_{\vec{q}^{\,\prime}} \, \hat{b}_{\vec{p}^{\,\prime}} \rangle
= \mathrm{Tr} \biggl[ \hat{\rho} \, \hat{b}^{\dagger}_{\vec{q}} \, \hat{b}^{\dagger}_{\vec{p}}\, \hat{b}_{\vec{q}^{\,\prime}} \, \hat{b}_{\vec{p}^{\,\prime}} \biggr],
\label{SC8}
\end{equation}
and similarly for all the other expectation values of the fields arising in Eq. (\ref{SC5}) upon insertion of Eq. (\ref{SC6}).
The density matrix appearing in Eq. (\ref{SC8}) can be the density matrix either of a pure state or of a mixed state and 
can be written, in general terms, as 
\begin{equation}
\hat{\rho} = \sum_{\{n\}} \, P_{\{n\}} \, | \{n\} \rangle \langle \{n\}|,\qquad \sum_{\{n\}} \, P_{\{n\}} =1.
\label{SC9}
\end{equation}
As already mentioned in section \ref{sec2} we are interested in the possibility that the initial state 
has a specified degree of second-order coherence. The simplest non-trivial situation is the one discussed in 
Eq. (\ref{sing7a}) and hereby generalized to the situation where $a\to a_{k}$ and $b\to b_{k}$ do depend upon the momentum 
(as appropriate in the case of many bosonic degrees of freedom). Thus the field theoretical generalization of the probability distribution implicitly mentioned in section \ref{sec2}, i.e. 
\begin{equation}
P_{\{n\}} = \prod_{\vec{k}} \frac{\Gamma(\zeta_{\vec{k}} + n_{\vec{k}})}{\Gamma(\zeta_{\vec{k}}) \Gamma(n_{\vec{k}} +1)} 
\biggl(\frac{\overline{n}_{\vec{k}}}{\overline{n}_{\vec{k}} + \zeta_{\vec{k}}}\biggr)^{n_{\vec{k}}} \, \biggl(\frac{\zeta_{\vec{k}}}{\overline{n}_{\vec{k}} + \zeta_{\vec{k}}}\biggr)^{\zeta_{\vec{k}}}.
\label{SC10}
\end{equation}
The evaluation of Eq. (\ref{SC8}) proceeds therefore by noticing that 
\begin{eqnarray}
&& \langle \hat{b}_{i}^{\dagger} \,  \hat{b}_{j}^{\dagger} \, \hat{b}_{k}\,  \hat{b}_{\ell} \rangle = 
\langle \hat{b}_{i}^{\dagger} \,  \hat{b}_{i}^{\dagger} \, \hat{b}_{i}\,  \hat{b}_{i}\rangle \delta_{i\,j} \,
\delta_{j\,k}\, \delta_{\ell\,k} +
\nonumber\\
&& \langle \hat{b}_{i}^{\dagger} \,  \hat{b}_{j}^{\dagger} \, \hat{b}_{i}\,  \hat{b}_{j} \rangle  \, \delta_{i\, k} \,
 \delta_{j\, \ell} [ 1 - \delta_{ij}] +  \langle \hat{b}_{i}^{\dagger} \,  \hat{b}_{j}^{\dagger} \, \hat{b}_{j}\,  \hat{b}_{i} \rangle  \, \delta_{i\, \ell} \,
 \delta_{j\, k} [ 1 - \delta_{ij}], 
\label{SC11}
\end{eqnarray}
where $\hat{b}_{i}$ and $\hat{b}_{j}^{\dagger}$ denote the annihilation and creation operators related to two generic momenta, 
i.e. for instance $\hat{b}_{\vec{q}}$ and $\hat{b}^{\dagger}_{\vec{p}}$; furthermore, following the same shorthand 
notation, $\delta_{i\, j}$ denotes the delta functions over the three-momenta (i.e. 
$ \delta_{\vec{q},\, \vec{p}}$).  In Eq. (\ref{SC11}) two different classes of terms appear: in the first class of terms (i.e. 
$\langle \hat{b}_{i}^{\dagger} \,  \hat{b}_{i}^{\dagger} \, \hat{b}_{i}\,  \hat{b}_{i}\rangle$) there are four different 
operators all acting on the {\em same} momentum; in the second class of terms (i.e. $ \langle \hat{b}_{i}^{\dagger} \,  \hat{b}_{j}^{\dagger} \, \hat{b}_{i}\,  \hat{b}_{j} \rangle$ and $\langle \hat{b}_{i}^{\dagger} \,  \hat{b}_{j}^{\dagger} \, \hat{b}_{j}\,  \hat{b}_{i} \rangle $) the momenta are paired two by two (as the corresponding deltas indicate).  Since 
the averages are to be computed using the initial density matrix characterized, in  general, by the 
statistical weights discussed above we have 
\begin{eqnarray}
 \langle \hat{b}_{i}^{\dagger} \,  \hat{b}_{j}^{\dagger} \, \hat{b}_{k}\,  \hat{b}_{\ell} \rangle &=& \sum_{m_{i}} m_{i} (m_{i} -1) P_{i}(m_{i}) 
\nonumber\\
\langle \hat{b}_{i}^{\dagger} \,  \hat{b}_{j}^{\dagger} \, \hat{b}_{i}\,  \hat{b}_{j} \rangle &=& \sum_{m_{i},\,\,m_{j}} m_{i}\,\, m_{j} \, P_{i\,j}(m_{i},m_{j}),
\nonumber\\
\langle \hat{b}_{i}^{\dagger} \,  \hat{b}_{j}^{\dagger} \, \hat{b}_{j}\,  \hat{b}_{i} \rangle &=& \sum_{m_{i},\,\,m_{j}} m_{i}\,\, m_{j} \, P_{i\,j}(m_{i},m_{j}).
\label{SC11a}
\end{eqnarray}
In Eq. (\ref{SC11a}) the following shorthand notations have been used
\begin{equation}
P_{i}(m_{i}) \equiv P_{k_{i}} (m_{k_{i}}), \qquad P_{ij}(m_{i},m_{j}) \equiv P_{k_{i}} (m_{k_{i}})\,P_{k_{j}} (m_{k_{j}}).
\label{SC11b}
\end{equation}
The second relation of Eq. (\ref{SC11b}) is indeed trivial in the light of the very definition of $P_{\{m\}}$:
\begin{equation}
P_{\{m\}} = P_{k_{1}}(m_{k_{1}})\,  P_{k_{2}}(m_{k_{2}})\, P_{k_{3}}(m_{k_{3}})\,....;
\label{SC12b}
\end{equation}
where the ellipses stand for the product over the various momenta; at the same time the explicit 
appearance of $P_{ij}$ is useful to trace the origin of the various terms.  Inserting Eq. (\ref{SC11a}) into Eq. (\ref{SC11}) 
the correlator becomes: 
\begin{eqnarray}
\langle \hat{b}_{i}^{\dagger} \,  \hat{b}_{j}^{\dagger} \, \hat{b}_{k}\,  \hat{b}_{\ell} \rangle &=&
\sum_{m_{i}} \biggl[ m_{i} (m_{i} -1) P_{i}(m_{i}) - 2 m_{i} \sum_{m_{j}} m_{j} P_{i\,j}(m_{i}, m_{j}) \biggr] \delta_{i\,j} \,
\delta_{j\,k}\, \delta_{\ell\,k}
\nonumber\\
&+& \sum_{m_{i},\,\,m_{j}} m_{i}\,\, m_{j} \, P_{i\,j}(m_{i},m_{j}) \biggl[ 
\delta_{i\,k} \delta_{j\, \ell} + \delta_{i\, \ell} \delta_{j \, k}\biggr].
\label{SC12a}
\end{eqnarray}
Finally, using Eqs. (\ref{SC11b}) and (\ref{SC12b}), Eq. (\ref{SC12a}) can be written as 
\begin{eqnarray}
\langle \hat{b}_{i}^{\dagger} \,  \hat{b}_{j}^{\dagger} \, \hat{b}_{k}\,  \hat{b}_{\ell} \rangle &=&
\sum_{m_{i}} \biggl[ m_{i} (m_{i} -1) P_{i}(m_{i}) - 2 m_{i} P_{i}(m_{i}) \sum_{m_{j}} m_{j} P_{j}(m_{j}) \biggr] \delta_{i\,j} \,
\delta_{j\,k}\, \delta_{\ell\,k}
\nonumber\\
&+& \sum_{m_{i}} m_{i} \, P_{i}(m_{i}) \sum_{m_{j}} m_{j} P_{j}(m_{j}) \biggl[ 
\delta_{i\,k} \delta_{j\, \ell} + \delta_{i\, \ell} \delta_{j \, k}\biggr].
\label{SC12}
\end{eqnarray}
For an explicit evaluation of the sums of Eq. (\ref{SC12}) it is useful to employ the probability generating function and the cumulant generating function 
\begin{equation}
 {\mathcal M} = \prod_{\vec{k}} {\mathcal M}_{\vec{k}}(s_{k}, \overline{n}_{\vec{k}}, \zeta_{\vec{k}}),\qquad 
 {\mathcal C} =  \prod_{\vec{k}} {\mathcal C}_{\vec{k}}(s_{k}, \overline{n}_{\vec{k}}, \zeta_{\vec{k}}),
 \label{SC13}
\end{equation} 
whose specific form, from Eq. (\ref{SC10}), becomes:
\begin{eqnarray}
&& {\mathcal M}_{\vec{k}}(s_{k}, \overline{n}_{\vec{k}}, \zeta_{\vec{k}}) = \frac{\zeta_{\vec{k}}^{\zeta_{\vec{k}}}}{[ \zeta_{\vec{k}} + (1 - s_{k}) \overline{n}_{\vec{k}}]^{\zeta_{\vec{k}}}},
\nonumber\\
&&  {\mathcal C}_{\vec{k}}(s_{k}, \overline{n}_{\vec{k}}, \zeta_{\vec{k}})= - \zeta_{\vec{k}} \, \ln{\biggl[ 1 + ( 1 - s_{k}) 
\frac{\overline{n}_{\vec{k}}}{\zeta_{\vec{k}}}\biggr]},
\label{SC14}
\end{eqnarray}
and coincides with the expression already derived in Eq. (\ref{sing10a}) in the case of a single degree of freedom.
The various sums of Eq. (\ref{SC12})  can be explicitly evaluated as combinations of the derivatives 
of the probability generating function. The final result can be expressed as 
\begin{equation}
\langle \hat{b}_{i}^{\dagger} \,  \hat{b}_{j}^{\dagger} \, \hat{b}_{k}\,  \hat{b}_{\ell} \rangle = 
\overline{n}_{i}^2 \biggl(\frac{1}{\zeta_{i}}-1\biggr) \delta_{ij} \, \delta_{jk} \, \delta_{k\ell} + \overline{n}_{i} \, \overline{n}_{j} \biggl[ \delta_{ik} \delta_{j \ell} + 
\delta_{i \ell} \delta_{j k} \biggr].
\label{SC15}
\end{equation}
By restoring, in Eq. (\ref{SC15}), the standard notations for the comoving three-momenta of the field 
we will have that:
\begin{equation}
\langle \hat{b}_{\vec{q}}^{\dagger} \,  \hat{b}_{\vec{p}}^{\dagger} \, \hat{b}_{\vec{q}^{\,\prime}}\,  \hat{b}_{\vec{p}^{\,\prime}} \rangle = 
\overline{n}_{q}^2 \biggl(\frac{1}{\zeta_{q}}-1\biggr) \delta_{\vec{q},\,\vec{p}} \, \delta_{\vec{p},\,\vec{q}^{\prime}} \, \delta_{ \vec{q}^{\,\prime},\,\vec{p}^{\,\prime}} 
+ \overline{n}_{q} \, \overline{n}_{p} \biggl[ \delta_{\vec{q},\, \vec{q}^{\,\prime}} \delta_{\vec{p},\, \vec{p}^{\,\prime}} + \delta_{\vec{q},\,\vec{p}^{\,\prime}} \delta_{\vec{p},\, 
\vec{q}^{\,\prime}} \biggr].
\label{SC15a}
\end{equation}
The same procedure leading to Eqs. (\ref{SC15}) and (\ref{SC15a}) can be used to compute, with the due differences, 
 all the other expectation values arising in the evaluation of the intensity correlation. In particular
\begin{eqnarray}
\langle \hat{b}_{i}\,  \hat{b}_{j} \, \hat{b}_{k}^{\dagger} \,  \hat{b}_{\ell}^{\dagger} \rangle &=& \overline{n}_{i}^2 \biggl(\frac{1}{\zeta_{i}}-1\biggr) \delta_{ij} \, \delta_{jk} \, \delta_{k\,\ell} + (\overline{n}_{i}+1) \, (\overline{n}_{j} +1) \biggl[ \delta_{i\,k} \delta_{j \,\ell} + 
\delta_{i \,\ell} \delta_{j\, k} \biggr],
\label{SC16a}\\
\langle \hat{b}_{i}\,  \hat{b}_{j}^{\dagger} \, \hat{b}_{k} \,  \hat{b}_{\ell}^{\dagger} \rangle &=& \overline{n}_{i}^2 \biggl(\frac{1}{\zeta_{i}}-1\biggr) \delta_{ij} \, \delta_{jk} \, \delta_{k\,\ell} + (\overline{n}_{i}+1) (\overline{n}_{k}+1) \delta_{i\,j} \delta_{k\,\ell} 
\nonumber\\
&+& \overline{n}_{k} (\overline{n}_{i} + 1) \delta_{i \,\ell} \delta_{j \, k},
\label{SC16b}\\
 \langle \hat{b}_{i}^{\dagger}\,  \hat{b}_{j} \, \hat{b}_{k}^{\dagger} \,  \hat{b}_{\ell} \rangle &=& \overline{n}_{i}^2 \biggl(\frac{1}{\zeta_{i}}-1\biggr) \delta_{ij} \, \delta_{jk} \, \delta_{k\,\ell} + \overline{n}_{i} \overline{n}_{k} \delta_{i\,j} \delta_{k\, \ell} + \overline{n}_{i} (\overline{n}_{k} + 1) \delta_{i\,\ell} \delta_{j \, k},
\label{SC16c}\\
\langle \hat{b}_{i}^{\dagger}\,  \hat{b}_{j} \, \hat{b}_{k} \,  \hat{b}_{\ell}^{\dagger} \rangle &=&  \overline{n}_{i}^2 \biggl(\frac{1}{\zeta_{i}}-1\biggr) \delta_{ij} \, \delta_{jk} \, \delta_{k\,\ell} + \overline{n}_{i} (\overline{n}_{\ell} + 1) \biggl[ \delta_{i\,j} \delta_{k\,\ell} + \delta_{i\, k}\delta_{\ell\, j}\biggr],
\label{SC16d}\\
 \langle \hat{b}_{i}\,  \hat{b}_{j}^{\dagger} \, \hat{b}_{k}^{\dagger} \,  \hat{b}_{\ell}\rangle &=&  \overline{n}_{i}^2 \biggl(\frac{1}{\zeta_{i}}-1\biggr) \delta_{ij} \, \delta_{jk} \, \delta_{k\,\ell} + \overline{n}_{\ell} (\overline{n}_{i} + 1) \biggl[ \delta_{i\,j} \delta_{k\,\ell} + \delta_{i\, k}\delta_{\ell\, j}\biggr].
\label{SC16e}
\end{eqnarray}
The physical role of the quantum correlations can be neatly understood by taking the Bose-Einstein limit in the correlators of Eqs. (\ref{SC15a}) and (\ref{SC16a})--(\ref{SC16e}). In the latter limit $\zeta_{i } \to 1$ for every mode of the field; Eq. (\ref{SC10}) turns into the standard thermal ensemble and the contributions 
of the terms of the type $\langle \hat{b}_{i}^{\dagger} \,\hat{b}_{i}^{\dagger}\, \hat{b}_{i} \, \hat{b}_{i} \rangle $ exactly cancels; the standard rules 
of evaluating correlators in thermal field theory is quickly recovered \cite{ford,kapusta}. It is useful, 
at this level, to take the limit $\zeta_{k} \to \infty$ uniformly for all modes of the field. From Eq. (\ref{SC14}) 
it can be immediately appreciated that, in the limit $\zeta_{k} \to \infty$, the probability generating function becomes 
${\mathcal M}_{k} \to \exp{[(s_{k}-1) \overline{n}_{k}]}$ which is exactly the generating function of the Poisson distribution. The Poisson distribution 
for each mode of the field is customarily associated with a multimode coherent state but this is not exactly our case: the situation 
described by Eq. (\ref{SC9}) and (\ref{SC10}) in the limit $\zeta_{k} \to \infty$ is the one of a {\em mixed} state with Poissonian 
distribution and this cannot be identified with a multimode coherent state since, in the latter case, the off-diagonal elements 
of the density matrix do not vanish (as it is the case for Eq. (\ref{SC10})). Finally, if the limit $\zeta_{k} \to 0$ is taken 
uniformly for all modes of the field, the probability distribution for each $k$-mode becomes logarithmic as it can be shown 
by using directly the recurrence relation characterizing the distribution of Eq. (\ref{SC10}).

We are then in condition of computing explicitly the intensity correlations in terms of an initial state characterized by the presence 
of quantum correlations reducing, in appropriate limits, to various statistical mixtures. The result can be written as 
\begin{eqnarray}
\langle \hat{{\mathcal I}}(\vec{x},\tau) \,  \hat{{\mathcal I}}(\vec{x} + \vec{r},\tau) \rangle &=&
 \int d\ln{k}\,\, {\mathcal G}^{(2)}_{\mathrm{v}}(k,\tau)\,\, [2 +  \cos{k r} j_{0}(k r)]
\nonumber\\
&+& \int d\ln{k}\,\, {\mathcal G}^{(2)}_{\mathrm{s}}(k,\tau) \,\,[1 + 2 j_{0}(k r)], 
\label{SC17}
\end{eqnarray}
where $j_{0}(k r)$ is the spherical Bessel function of zeroth order; the two functions ${\mathcal G}^{(2)}_{\mathrm{v}}(k,\tau)$
and ${\mathcal G}^{(2)}_{\mathrm{s}}(k,\tau)$ denote, respectively, the bulk (volume) and boundary (surface) contributions
\begin{eqnarray}
&& {\mathcal G}^{(2)}_{\mathrm{v}}(k,\tau) = \frac{ k\,\, \overline{n}^2_{k}(\tau)}{4 \pi^2 V} \,\,\biggl( \frac{1}{\zeta_{k}} - 1\biggr) \biggl[ 2 \overline{N}_{k}(\tau) + 1 - 
2 \sqrt{\overline{N}_{k} (\overline{N}_{k} +1) } \cos{\alpha_{k}}\biggr]^2,
\label{SC18}\\
&&  {\mathcal G}^{(2)}_{\mathrm{s}}(k,\tau) = \int \frac{d^{3} q}{64\pi^5} \, \frac{k^3}{q | \vec{k} - \vec{q}|} {\mathcal F}(q,\tau) {\mathcal F}(|\vec{k} - \vec{q}|, \tau) 
\label{SC19}
\end{eqnarray}
with $r = |\vec{x} - \vec{y}|$  and 
\begin{equation}
{\mathcal F}(q,\tau) = [2 \,\,\overline{n}_{q}(\tau) +1 ] \,\,\biggl\{ 2 \overline{N}_{q}(\tau) + 1 - 2 \sqrt{\overline{N}_{q}(\tau) \,[\overline{N}_{q}(\tau) +1]} \, \cos{\alpha_{q}(\tau)}\biggr\}.
\label{SC20}
\end{equation}
We shall also assume, from now on, that $\overline{n}_{\vec{k}} = \overline{n}_{k}$ and that 
$\zeta_{\vec{k}} = \zeta_{k}$ with $|\vec{k}| = k$.
The denominator appearing in the degree of second-order coherence is 
\begin{eqnarray}
\langle \hat{{\mathcal I}}(\vec{x}, \tau)\rangle\langle \hat{{\mathcal I}}(\vec{y},\tau) \rangle
&=& \biggl| \int \, d\ln{k} \,{\mathcal G}^{(1)}(k,\tau) \, \biggl|^2,
\label{SC21}\\
{\mathcal G}^{(1)}(k,\tau) &=& \frac{k^2}{4\pi^2} ( 2 \overline{n}_{k} + 1) \biggl[2 \overline{N}_{k} + 1 - 2 \sqrt{\overline{N}_{k}(\overline{N}_{k} +1)} \cos{\alpha_{k}}\biggr],
\end{eqnarray}
where the superscript reminds that we are dealing here with the square of the first-order correlation discussed 
in section \ref{sec3}.
The normally ordered definition of the degree of second-order coherence introduced in Eq. (\ref{SC3}) leads to a result which is somehow 
similar to the one obtained in the case of Eq. (\ref{SC1}). For future comparison the normal ordered intensity correlation 
can be written, in explicit terms, as 
\begin{eqnarray}
\langle :\hat{\Phi}^{(-)}(\vec{x}, \tau) \, \hat{\Phi}^{(-)}(\vec{y}, \tau)
\hat{\Phi}^{(+)}(\vec{y}, \tau) \, \hat{\Phi}^{(+)}(\vec{x}, \tau):\rangle &=&
 \int d\ln{k}\,\overline{{\mathcal G}}^{(2)}_{\mathrm{v}}(k,\tau)  
 \nonumber\\
&+& \int d\ln{k}\, \overline{{\mathcal G}}^{(2)}_{\mathrm{s}}(k,\tau) [ 1 + j_{0}(k r)],
\label{SC22}
\end{eqnarray}
where now 
\begin{eqnarray}
&& \overline{{\mathcal G}}^{(2)}_{\mathrm{v}}(k,\tau)  = \frac{ k \, \overline{n}_{k}^2 }{4 \pi^2 \, V} \biggl( \frac{1}{\zeta_{k}} - 1 \biggr) \biggl[ 2 \overline{N}_{k}  + 1 - 
2 \sqrt{\overline{N}_{k} (\overline{N}_{k} + 1)} \cos{\alpha_{k}} \biggr]^2 
\label{SC23}\\
&&  \overline{{\mathcal G}}^{(2)}_{\mathrm{s}}(k,\tau) = \int \frac{d^{3} q}{64\pi^{5} }\, \frac{k^3}{q |\vec{k} - \vec{q}|} \, \overline{{\mathcal F}}(q,\tau) \, \overline{{\mathcal F}}(|\vec{q} - \vec{k}|,\tau),
\label{SC24}\\
&& \overline{{\mathcal F}}(q,\tau) = \overline{n}_{q} \biggl[ 2 \overline{N}_{q} + 1 - 2 \sqrt{\overline{N}_{q} (\overline{N}_{q} +1)} \, \cos{\alpha_{q}} \biggr].
\label{SC25}
\end{eqnarray}
Of course in case the normal ordered expression is used, also the denominator of Eq. (\ref{SC3}) 
must be ordered and computed accordingly. The result is
\begin{eqnarray}
&& \langle \hat{\Phi}^{(-)}(\vec{x},\tau) \, \hat{\Phi}^{(+)}(\vec{x},\tau) \rangle \langle  \hat{\Phi}^{(-)}(\vec{y},\tau) \, \hat{\Phi}^{(+)}(\vec{y},\tau)\rangle =
\biggl| \int d\ln{k}\,\overline{{\mathcal G}}^{(1)}(k, \tau)\,\biggr|^2.
\label{SC26}\\
&& \overline{{\mathcal G}}^{(1)}(k, \tau) =  \frac{k^2}{2\pi^2}\, \overline{n}_{k}\, \biggl[ 2 \overline{N}_{k}  + 1 - 2 \sqrt{\overline{N}_{k}(\overline{N}_{k} +1)}\, \cos{\alpha_{k}}\biggr].
\label{SC27}
\end{eqnarray}
There are two main differences between the normal ordered correlators and the non-normal ordered ones: a 
numerical factor in Eq. (\ref{SC22}) and the ubiquitous presence of $ \overline{n}_{k}$ instead of $(2 \overline{n}_{k} +1)$.
The expressions of Eqs. (\ref{SC23}) and (\ref{SC24}) vanish in the limit $\overline{n}_{k} \to 0$ while in the case of Eqs. 
(\ref{SC18}) and (\ref{SC19}) the same limit does not vanish.  Furthermore, the degree of second-order coherence inherits a volume dependence which is directly linked to the existence 
of an initial state with a non vanishing degree of second-order correlation. 
To conclude the present section it is appropriate to mention that the averages over the initial state can also be 
conducted by making appropriate use of the phase-space functional previously discussed (see Eqs. (\ref{F9A})--(\ref{F10A})).
In particular, the multiparticle states defined by the density matrix of Eqs. (\ref{SC9}) and (\ref{SC10}) leads to a $P$-representation
given by
\begin{equation}
P(\{\gamma\}) = \prod_{\vec{k}} \frac{\zeta_{k}^{\zeta_{k}}}{\overline{n}_{k}\, \Gamma(\zeta_{k})} \, |\gamma_{k}|^{ - 2 ( 1 - \zeta_{k})} 
\, e^{ - |\gamma_{k}|^2 \zeta_{k} /\overline{n}_{k}}.
\label{SC28}
\end{equation}
The $P$-representation of Eq. (\ref{SC28}) has interesting limits. In particular for 
$\zeta_{k} \to 1$ the $P$-representation of Eq. (\ref{F11A}) is reproduced.  The representation (\ref{SC28}) has been first discussed in \cite{matsuo} in the purely 
quantum mechanical case.

\renewcommand{\theequation}{6.\arabic{equation}}
\setcounter{equation}{0}
\section{Degree of second-order coherence}
\label{sec6}
\subsection{Basic considerations}
The degree of second-order coherence 
discussed in section \ref{sec5} will now be evaluated explicitly in various 
potentially interesting situations. 
The specific values of the cosmological parameters determined using the WMAP 7yr data alone 
in the light of the vanilla $\Lambda$CDM scenario \cite{WMAP7a,WMAP7b} are \footnote{Note that 
$\varepsilon_{\mathrm{re}}$ denotes the optical depth at reionization and has nothing 
to do with the slow-roll parameter $\epsilon$ which will be introduced in a moment.}: 
\begin{equation}
( \Omega_{\mathrm{b}}, \, \Omega_{\mathrm{c}}, \Omega_{\mathrm{de}},\, h_{0},\,n_{\mathrm{s}},\, \varepsilon_{\mathrm{re}}) \equiv 
(0.0449,\, 0.222,\, 0.734,\,0.710,\, 0.963,\,0.088),
\label{Par1}
\end{equation}
where $\Omega_{X}$ denotes the present critical fraction of the corresponding 
species (i.e., respectively, baryons, CDM particles, dark energy); $n_{\mathrm{s}}$ denotes the scalar spectral index (see also 
Eqs. (\ref{Par4}), (\ref{DS4}) and (\ref{DS5}) hereunder). If a tensor component is allowed in the analysis 
of the WMAP 7yr data alone the relevant cosmological parameters are determined to be \cite{WMAP7a,WMAP7b} 
\begin{equation}
( \Omega_{\mathrm{b}}, \, \Omega_{\mathrm{c}}, \Omega_{\mathrm{de}},\, h_{0},\,n_{\mathrm{s}},\, \varepsilon_{\mathrm{re}}) \equiv 
(0.0430,\, 0.200,\, 0.757,\,0.735,\, 0.982,\,0.091).
\label{Par2}
\end{equation}
In the case of Eq. (\ref{Par1}) the amplitude of the scalar modes is ${\mathcal A}_{{\mathcal R}} = 
(2.43 \pm 0.11) \times 10^{-9}$ while in the case of Eq. (\ref{Par2}) the corresponding values of ${\mathcal A}_{{\mathcal R}}$ and of $r_{\mathrm{t}}$ are given by 
\begin{equation}
{\mathcal A}_{{\mathcal R}} = (2.28 \pm 0.15)\times 10^{-9},\qquad r_{\mathrm{t}} < 0.36 
\label{Par3}
\end{equation}
to $95$ \% confidence level. The experimental parametrization of the scalar and tensor power spectra is \cite{WMAP7a,WMAP7b}
\begin{equation}
{\mathcal P}_{{\mathcal R}}(k) = {\mathcal A}_{{\mathcal R}}   \biggl(\frac{k}{k_{\mathrm{p}}}\biggr)^{n_{\mathrm{s}}-1},\qquad {\mathcal P}_{\mathrm{t}}(k) = {\mathcal A}_{\mathrm{t}} \biggl(\frac{k}{k_{\mathrm{p}}}\biggr)^{n_{\mathrm{t}}}, 
\label{Par4}
\end{equation}
where $n_{\mathrm{t}}$, $n_{\mathrm{s}}$ are, respectively, 
the tensor spectral index, and  
$k_{\mathrm{p}} = 0.002 \,\, \mathrm{Mpc}^{-1}$ is the pivot scale;
 $r_{\mathrm{t}} = {\mathcal A}_{\mathrm{t}}/{\mathcal A}_{{\mathcal R}}$ 
denotes the ratio between the tensor and the scalar power spectrum at $k_{\mathrm{p}}$. The qualitative features 
of the effects discussed here do not change if, for instance, one 
would endorse the parameters drawn from the minimal tensor extension 
of the $\Lambda$CDM paradigm and compared not to the WMAP 7yr 
data release but rather with the WMAP 3yr data release \cite{WMAP3a,WMAP3b}, implying, for instance, ${\mathcal A}_{{\mathcal R}} = 2.1^{+2.2}_{-2.3}\times 10^{-9}$, 
$n_{\mathrm{s}} =0.984$ and  $r_{\mathrm{t}} < 0.65$ (95 \% confidence level).

The numeric values reported in Eqs. (\ref{Par1}), (\ref{Par2}) and (\ref{Par3}) 
determine bounds on the slow roll parameters appearing directly 
in the evaluation of the degree of second-order coherence. In particular,
within the present notations, the slow-roll parameters are defined as
where, as usual,  
\begin{eqnarray}
&& \epsilon = - \frac{\dot{H}}{H^2} = \frac{\overline{M}_{\mathrm{P}}^2}{2} \biggl(\frac{V_{,\varphi}}{V}\biggr)^2,
\nonumber\\
&& \eta = \frac{\ddot{\varphi}}{H \dot{\varphi}} = \epsilon - \overline{\eta},\qquad \overline{\eta} = 
\overline{M}_{\mathrm{P}}^2 \frac{V_{,\varphi\varphi}}{V},
\label{DS4}
\end{eqnarray}
and $\overline{M}_{\mathrm{P}} = M_{\mathrm{P}}/\sqrt{8 \pi}$ is the reduced 
Planck mass, $H$ is the Hubble parameter, $V$ is the inflaton 
potential and the overdot denotes a derivation with respect to the cosmic time\footnote{Recall that, as usual, the relation 
between cosmic and conformal time parametrization is given by $a(\tau) \, d\tau = dt$.} coordinate $t$. To lowest order in the slow-roll expansion we also have that 
\begin{equation}
n_{\mathrm{t}} = - 2 \epsilon, \qquad n_{\mathrm{s}} = 1 - 6 \epsilon + 2 \overline{\eta}, 
\qquad r_{\mathrm{t}} = 16 \epsilon = - 8 n_{\mathrm{t}}.
\label{DS5}
\end{equation}
The mean number of particles per Fourier mode can be obtained from 
Eqs. (\ref{EQQ21}). While $\overline{n}_{k}$  and $\zeta_{k}$ 
are a property of the initial state, the quantities 
$\overline{N}_{k}$ and $\alpha_{k}$ can be computed explicitly, for instance, 
in the case of a single-field inflationary model. 
To get the explicit solutions either in terms of Eqs. (\ref{EQQ21})  or in terms 
of  Eqs. (\ref{H2b})--(\ref{H2c}) background fields need to be expressed in terms 
of the conformal time coordinate and in terms of the slow-roll parameters of Eq. (\ref{DS4}):
\begin{eqnarray}
\partial_{\tau} \ln{a} &=& a H = - \frac{1}{\tau ( 1 - \epsilon)},
\label{Par5a}\\
\partial_{\tau}\ln{z} &=& - \frac{1 + \epsilon + \eta}{\tau ( 1 - \epsilon)},
\label{Par5}\\
\partial^2_{\tau} \ln{a} + (\partial_{\tau} \ln{a})^2  &=& \frac{ 2 - \epsilon}{\tau^2 ( 1 -\epsilon)^2},
\label{Par6}\\
\partial^2_{\tau} \ln{z} + (\partial_{\tau} \ln{z})^2 &=& \frac{2 + 2 \epsilon + 3 \eta + \epsilon\eta + \eta^2}{( 1 - \epsilon)^2 \tau^2}.
\label{Par7}
\end{eqnarray}
Recalling that $\overline{N}_{k} = 
\sinh^2{r_{k}}$ and denoting with $\overline{N}^{(\mathrm{p})}$ 
and $\overline{N}^{(\mathrm{g})}$ the average number of phonons and gravitons 
the final result turns out to be:
\begin{eqnarray}
\overline{N}_{k}^{(\mathrm{p})} + \frac{1}{2} &=&  \frac{\pi}{4} (- k \tau) \biggl[ 
\bigl| H_{\beta}^{(1)}( - k \tau) \bigr|^2 + \bigl| H_{\beta-1}^{(1)}( - k \tau) \bigr|^2\biggr],
\label{DS1}\\
\overline{N}_{k}^{(\mathrm{g})} + \frac{1}{2} &=&  \frac{\pi}{4} (- k \tau) \biggl[ 
\bigl| H_{\alpha}^{(1)}( - k \tau) \bigr|^2 + \bigl| H_{\alpha-1}^{(1)}( - k \tau) \bigr|^2\biggr],
\label{DS2}
\end{eqnarray}
where $H_{\gamma}^{(1)}(z)$ is the Hankel function of the first kind of index $\gamma$ and argument $z$  \cite{abr,tric}. In Eqs. (\ref{DS1}) and (\ref{DS2}) the indexes of the Hankel functions depend upon the slow-roll parameters as
\begin{equation}
\alpha = \frac{3 -\epsilon}{2 ( 1 - \epsilon)}, \qquad \beta = \frac{3 + \epsilon + 2 \eta}{2 ( 1 - \epsilon)}.
\label{DS3}
\end{equation}
Using Eq. (\ref{DS5}) 
$\alpha$ and $\beta$ can be written directly in terms of the scalar and tensor 
spectral indices
\begin{equation}
\alpha = \frac{ 6 + n_{\mathrm{t}}}{2(2 + n_{\mathrm{t}})},\qquad 
\beta = \frac{8 + 3 n_{\mathrm{t}} - 2 n_{\mathrm{s}}}{2(2  + n_{\mathrm{t}})}.
\label{DS6}
\end{equation}
In the range of parameters mentioned in Eqs. (\ref{Par1}), (\ref{Par2}) and (\ref{Par3}) 
the indexes $\alpha >1$ and $\beta>1$ are always positive definite and, moreover, 
\begin{eqnarray}
&& ( 3 - 2 \alpha) = - \frac{2 \epsilon}{1 - \epsilon} < 0, \qquad ( 3 - 2 \beta) = - \frac{4 \epsilon + 2 \eta}{( 1 -\epsilon)},
\nonumber\\
&& ( 3 - 4 \alpha) = - \frac{3+ \epsilon}{1 - \epsilon} < 0, \qquad ( 3 - 4 \beta) = - \frac{3 +9 \epsilon + 4 \eta}{( 1 -\epsilon)}.
\label{DS22}
\end{eqnarray}
Equations (\ref{DS1}) and (\ref{DS2}) 
can be expanded in the limits $|k\tau| \gg 1$ and $|k\tau |\ll 1$ 
holding, respectively, when the corresponding wavelengths are either 
shorter or larger than the Hubble radius. In the limit $|k\tau | \ll 1$,
Eqs. (\ref{DS1}) and (\ref{DS2}) become:
\begin{eqnarray}
\overline{N}_{k}^{(\mathrm{p})} + \frac{1}{2} &=&  \frac{\Gamma^2(\beta)}{2 \pi} \biggl(- 
\frac{k \tau}{2} \biggr)^{ 1 - 2 \beta} \biggl[ 1 + \frac{|k\tau|^2}{4 (\beta-1)^2}\biggr],
\label{DS7}\\
\overline{N}_{k}^{(\mathrm{g})} + \frac{1}{2} &=&   \frac{\Gamma^2(\alpha)}{2 \pi} \biggl(- 
\frac{k \tau}{2} \biggr)^{ 1 - 2 \beta} \biggl[ 1 + \frac{|k\tau|^2}{4 (\alpha-1)^2}\biggr].
\label{DS8}
\end{eqnarray} 
In the limit of short wavelengths the particles are all inside the Hubble radius 
and, consequently, 
\begin{equation}
\overline{N}^{(\mathrm{p})} = \overline{N}^{(\mathrm{g})} \to  \frac{1}{2},
\label{DS9}
\end{equation}
where the factor $1/2$ confirms, a posteriori, the correctness of all the normalizations and implies that, in the 
vacuum, there is half a quantum per Fourier mode.
The degree of second-order coherence is also determined by a phase (see, e.g. Eq. (\ref{SC20})) whose explicit form for phonons and gravitons can be written as 
\begin{eqnarray}
\cos{\alpha^{(\mathrm{p})}_{k}} &=& \frac{\pi}{4} (- k \tau)  \frac{\bigl|H_{\beta-1}^{(1)}(-k\tau)\bigr|^2 - \bigl|H_{\beta}^{(1)}(-k\tau)\bigr|^2}{\sqrt{N_{k}^{(\mathrm{p})} (N_{k}^{(\mathrm{p})}+1)}},
\label{DS10}\\
\cos{\alpha^{(\mathrm{g})}_{k}} &=& \frac{\pi}{4} (- k \tau)  \frac{\bigl|H_{\alpha-1}^{(1)}(-k\tau)\bigr|^2 - \bigl|H_{\alpha}^{(1)}(-k\tau)\bigr|^2}{\sqrt{N_{k}^{(\mathrm{g})} (N_{k}^{(\mathrm{g})}+1)}},
\label{DS11}
\end{eqnarray}
where, consistently with the notation of Eqs. (\ref{DS1}) and (\ref{DS2}), the superscripts refer, respectively, to the case of 
the gravitons and of the phonons.
In the limits $|k\tau| \ll 1$ and $|k\tau| \gg 1 $, Eqs. (\ref{DS10}) and (\ref{DS11}) 
lead, respectively, to
\begin{equation}
\lim_{|k\tau| \to 0 } \cos{\alpha_{k}^{(\mathrm{s,\,t})}} \to -1,\qquad 
\lim_{|k\tau| \to \infty } \cos{\alpha_{k}^{(\mathrm{s,\,t})}} \to 0,
\label{DS12}
\end{equation}
where the superscripts simply mean that the mentioned limits hold separately for the scalar and for the 
tensor modes. 

\subsection{Initial vacuum state}
The relevant physical limit of the degree of second-order 
coherence can be expressed as follows:
\begin{equation}
\lim_{|k\tau| \to 0 ,\,\,\, |k\,r|\to 0} g^{(2)}(\vec{x}, \vec{x} + \vec{r};\tau).
\label{DS13}
\end{equation}
The limits appearing in Eq. (\ref{DS13}) imply that the degree of second-order coherence 
is evaluated at coincidental spatial points and when the comoving momenta are much shorter either than
 $\partial_{\tau} \ln{a}$ (in the case of the gravitons) or than $\partial_{\tau} \ln{z}$ (in the case of the phonons). 
As in the case of the limit discussed in Eq. (\ref{F3A}) the mathematical definition of the limit 
only involves $r$ since the degree of second-order coherence is always defined as the ratio 
of integrals over the comoving three-momentum. At the same time the limit $r\to 0$ is physically 
realized for typical wavenumbers  $|k\,r|\to 0$ and $|k\tau| \to 0$ and this defines, according to Eq. 
(\ref{DS13}) the normalized degree of second-order coherence in the asymptotic limit 
of typical scales (or wavelengths) larger than the Hubble radius. This kind of procedure has some analogy 
with quantum optics where often the results of the momentum integrations in the  numerator and in the denominator 
of the degree of second-order coherence simplify in the limit $r \to 0$ and lead to a degree of second-order 
coherence which only depends on time (see, e.g., \cite{loudon,mandel}). 
 
It is useful to remark that, a posteriori, the order of the limits appearing Eq. (\ref{DS13}) is not essential: the same results can be obtained by changing the order in which the limits are taken. 
From the explicit expressions of Eq. (\ref{SC17}) the degree of second-order coherence can be written as 
\begin{equation}
\lim_{k r \to 0}  g^{(2)}(r, \tau) = 3 \frac{\int d\,\ln{k} \, [ {\mathcal G}^{(2)}_{\mathrm{v}}(k,\tau) + 
{\mathcal G}^{(2)}_{\mathrm{s}}(k,\tau)]}{\bigl| \int d\,\ln{k} \, {\mathcal G}^{(1)}(k,\tau)\bigr|^2}, 
\label{DS14}
\end{equation}
recalling that $j_{0}(z) \to 1$ for $z \to 0$. Using all the results derived in the present section 
(and, in particular, Eqs. (\ref{DS7}), (\ref{DS8}) and (\ref{DS12})),  also the limit $|k\tau| \to 0$ can be 
taken with the result that\footnote{The result will be given in the case of the gravitons. The results 
for the phonons can be obtained from the ones of the gravitons by replacing $\alpha \to \beta$.}
\begin{eqnarray}
\lim_{|k\tau| \to 0 ,\,\,\, |k\,r|\to 0} g^{(2)}(r,\tau) &=& {\mathcal A}_{\mathrm{v}}(\alpha) 
 + {\mathcal B}_{\mathrm{s}}(\alpha),
\label{DS15}\\
{\mathcal A}_{\mathrm{v}}(\alpha) &=& 
 \frac{3\pi^2}{V} \frac{\int d\ln{k} \, \overline{n}_{k}^2 ( 1 - \zeta_{k})/\zeta_{k}\,\,k^{3 - 4\alpha}}{\bigl| \int d\ln{k} ( 2 \overline{n}_{k} + 1) k^{ 3 - 2\alpha}\bigr|^2},
\label{DS15A}\\
{\mathcal B}_{\mathrm{s}}(\alpha) &=&\frac{3}{2} \frac{\int d\ln{k} k^{3} \int d\ln{q} q^{3- 2\alpha} ( 2 \overline{n}_{q} +1) {\mathcal Y}(|\vec{k}-\vec{q}|)}{\bigl| \int d\ln{k} ( 2 \overline{n}_{k} + 1) k^{ 3 - 2\alpha}\bigr|^2},
\label{DS15B}
\end{eqnarray}
where 
\begin{equation}
{\mathcal Y}(|\vec{k}-\vec{q}|) = \int_{-1}^{1} d x \frac{2 \overline{n}_{|\vec{k} - \vec{q}|}+1}{|\vec{k} - \vec{q}|^{2\alpha}},\qquad 
|\vec{k} - \vec{q}| = \sqrt{ q^2 + k^2 - 2\, q\, x}\,.
\label{DS16}
\end{equation}
The initial vacuum state is recovered when $\overline{n}_{k} \to 0$  implying that the particle content 
of the initial state vanishes. In the latter limit, in particular, we have 
\begin{eqnarray}
&& \lim_{\overline{n}_{k} \to 0} {\mathcal A}_{\mathrm{v}}(\alpha) = 0,
\label{DS16A}\\
&&  \lim_{\overline{n}_{k} \to 0} {\mathcal B}_{\mathrm{s}}(\alpha) = \frac{3}{2} \frac{\int d\ln{k} k^{3} \int d\ln{q} q^{3- 2\alpha} {\mathcal Y}(|\vec{k}-\vec{q}|)}{\bigl| \int d\ln{k}  k^{ 3 - 2\alpha}\bigr|^2};
\label{DS16B}
\end{eqnarray}
the same result holds, as previously remarked, by exchanging $\alpha$ withe $\beta$.
By introducing the rescaling  $y = q/k$ the explicit form of Eq. (\ref{DS16B}) becomes
\begin{equation}
{\mathcal B}_{\mathrm{s}}(\alpha)= 
\frac{3}{2} \frac{\int d \ln{k}\, k^{ 3 - 2 \alpha} \, \int d\ln{q} \, 
q^{3 - 2 \alpha}\, \int_{-1}^{1} \frac{d x}{(1 + y^2 - 2 y x)^{\alpha}}}{\bigl| \int d\ln{k}  k^{ 3 - 2\alpha}\bigr|^2}.
\label{DS19}
\end{equation}
After integration over the $x$ variable, the numerator of Eq. (\ref{DS19}) becomes 
\begin{eqnarray}
&& \int d \ln{k}\, k^{ 3 - 2 \alpha} \, \int d\ln{q} \, 
q^{3 - 2 \alpha}\, \int_{-1}^{1} \frac{d x}{(1 + y^2 - 2 y x)^{\alpha}} = 
\nonumber\\
&& \frac{1}{2 (1 -\alpha)} \int \, d\ln{k}\,\, k^{6 - 4 \alpha} 
\int d y\,\, y^{1 - 2 \alpha} \biggl[ |1 + y|^{2 ( 1 - \alpha)} - | 1 - y|^{ 2 ( 1 - \alpha)}\biggr].
\label{DS20}
\end{eqnarray}
The value of the integral over $y= q/k$ can be estimated as 
\begin{equation}
4( 1 -\alpha) \biggl\{ \frac{1}{3 - 2 \alpha} \biggl[ 1 - y_{\mathrm{min}}^{3 - 2 \alpha}\biggr] + 
\frac{1}{3 - 4 \alpha}\biggl[ y_{\mathrm{max}}^{ 3 - 4 \alpha} -1\biggr]\biggr\}.
\label{DS21}
\end{equation}
The limit $y_{\mathrm{max}} \to \infty$ can be taken\footnote{It is worth stressing that the negative 
sign of the combinations $(3 - 4 \alpha)$ and $(3 - 4 \beta)$ is a direct consequence of the determinations 
of cosmological parameters reported in Eqs. (\ref{Par1}), (\ref{Par2}) and (\ref{Par3}).}   since $(3 - 4\alpha) <0$ 
(and $( 3 - 4 \beta) < 0$). Conversely the value of $y_{\mathrm{min}} = q_{0}/k$ 
forbids taking the limit $y_{\mathrm{min}} \to 0$ since 
$ (3 - 4 \alpha) $ and $ ( 3 - 4\beta)$ are both negative as established 
in Eq. (\ref{DS22}).  Since the term containing $y_{\mathrm{min}}$ dominates the integral of  Eq. (\ref{DS20})
can be estimated up to sub-leading corrections. Integrating over $k$ and going then back to Eq. (\ref{DS15}) we have
\begin{equation}
 \lim_{|k\tau| \to 0 ,\,\,\, |k\,r|\to 0} g^{(2)}(r,\tau)  = 3.
\label{DS24}
\end{equation}
The value given by Eq. (\ref{DS24}) coincides with the value obtained, in the case of a single 
degree of freedom (and for the squeezed vacuum state) as discussed in section \ref{sec2}. As mentioned in section \ref{sec2} the normal ordering in the operator
(or its absence) does not affect the degree of second-order coherence 
as long as the average number of particles per field mode is much larger 
than one. This is, a posteriori, exactly the physical limit of Eqs. (\ref{DS1})--(\ref{DS2}).
In connection with Eq. (\ref{DS24}) two final remark are in order. As mentioned after 
Eq. (\ref{DS13}) the limits appearing in Eq. (\ref{DS24}) signify that 
the result holds for typical comoving three-momenta 
larger than the Hubble radius; the strict mathematical limit would 
instead stipulate that $ r \to 0$.  The limit defined in Eqs. (\ref{DS13}) and (\ref{DS24}) 
demands that the integrals appearing in the numerator and in the denominator 
of the degree of second-order coherence are evaluated for typical wavelengths 
larger than the Hubble radius.

\subsection{Classical stochastic variables}
The result obtained in Eq. (\ref{DS24}) implies that the relic gravitons 
and relic phonons are highly bunched and their statistics is super-Poissonian.
The large-scale curvature fluctuations might also be classical stochastic 
variables. In the latter case, however, the degree of second-order coherence 
in the zero time-delay limit will typically be between $1$ and $2$, i.e.
 $1 \leq g^{(2)}  \leq 2$.  In this case the averages appearing in $g^{(2)}$ will have to 
 be interpreted as stochastic averages. 
The lower limit (i.e. $g^{(2)}=1$) is easy to justify. Suppose, indeed, that 
the relic phonons (or gravitons) are just produced independently. This means 
that the intensity ${\mathcal I}$ can be viewed as a classical (discrete) 
variable characterized by a Poissonian probability distribution.  In this case 
$\langle {\mathcal I}^{\,r} \rangle = \langle {\mathcal I} \rangle^{r}$  and, 
consequently, $g^{(2)} =1$. In is also possible to conceive a completely 
classical situation where the source is chaotic in such a way 
that $\langle {\mathcal I}^{\,r} \rangle = \, r!\, \langle {\mathcal I} \rangle^{r}$. 
In the latter case the probability distribution for the (possibly time- or space-dependent) 
intensity can be written as $P({\mathcal I}) = \overline{{\mathcal I}}^{-1} \, 
\exp{[- {\mathcal I}/\overline{{\mathcal I}}]}$, where $ \overline{{\mathcal I}} = 
\langle {\mathcal I} \rangle$.  

\subsection{Finite volume effects} 
As a next step of complication it is useful to analyze the case where the number 
of particles in each Fourier mode is the same but second-order correlations are allowed in the 
initial state, i.e.
\begin{equation}
\overline{n}_{k} = \overline{n},\qquad \zeta_{k} = \zeta.
\label{DS25}
\end{equation}
Following the same steps outlined in the vacuum situation, the analog of the limit 
given in Eq. (\ref{DS24}) is 
\begin{equation}
 \lim_{|k\tau| \to 0 ,\,\,\, |k\,r|\to 0} g^{(2)}(r,\tau)  = 3 + \frac{3\pi^2}{k_{0}^ 3 V} \frac{(1 - \zeta)}{\zeta} \frac{\overline{n}^2}{( 2 \overline{n}+	1)^2} \frac{ (3 - 2\alpha)^2}{4\alpha -3}.
\label{DS25A}
\end{equation}
Concerning this expression few comments are in order:
\begin{itemize}
\item{} the first term (i.e. $3$) remains also in the limit $\overline{n} \to 0$ (i.e. in the vacuum case);
\item{} the second term goes to zero in the infinite volume limit (which is the one which must be enforced) and also 
in the case of a Bose-Einstein distribution (i.e. $\zeta \to 1$); 
\item{} the numerical factors in the second term depend upon the slow-roll parameter.
\end{itemize}  
The result of Eq. (\ref{DS25A}) shows, in a specific example, 
 that second-order correlation effects possibly present in the initial state vanish when the volume goes 
 to infinity and the average multiplicity goes to infinity (while their ratio is kept fixed).  The implications 
 of this result are interesting per se but a closer scrutiny goes beyond the aims of this script. 
 
 \subsection{Bose-Einstein occupation number}
As suggested by the general equations derived in section \ref{sec6}, the degree of second-order coherence is sensitive to the overall duration of the inflationary 
phase even in the case $\zeta_{k} = 1$ where, by definition, volume 
effects are absent and the generalized statistical ensemble of Eq. (\ref{SC10}) 
reduces to the thermal (or chaotic) one (see Eq. (\ref{F7A})). Resorting to the quantum mechanical 
analogy discussed in section \ref{sec2} it would be tempting to conclude on the basis 
of Eq. (\ref{sing10}), that the degree of second-order coherence should still equal 
$3$. Indeed, the limit $\overline{N} \simeq \overline{n} \gg 1$ 
and $\zeta \to 1$ implies, from Eq. (\ref{sing10}), that $\overline{g}^{(2)} \to 3$.
This conclusion is incorrect as long as, in the realistic field theoretical case, 
$\overline{N}_{k} \neq \overline{n}_{k}$; more specifically  
\begin{equation}
\overline{n}_{k} = \frac{1}{e^{k/k_{\mathrm{T}}} -1},\qquad \zeta_{k} = 1,
\label{DS17}
\end{equation}
where $k_{\mathrm{T}}$ is the (comoving) thermal momentum whose 
explicit value can be usefully expressed in Hubble units (i.e. units 
of the Hubble rate $H_0$) as
\begin{equation}
\frac{k_{\mathrm{T}}}{H_{0}} =  e^{N_{\mathrm{max}} - N_{\mathrm{tot}}}\, 
\biggl(\frac{T}{H}\biggr)\,\biggl(\frac{{\mathcal A}_{{\mathcal R}}}{2.43 \times 10^{-9}}\biggr)^{1/4} \, \biggl(\frac{h_{0}^2 \Omega_{\mathrm{R}0}}{4.15 \times 10^{-9}}\biggr)^{1/4} \biggl(\frac{0.7}{h_{0}}\biggr).
\label{DS18}
\end{equation}
In Eq. (\ref{DS18}) the comoving value of the thermal momentum\footnote{As already mentioned we shall normalize
to $1$ the present value of the scale factor $a_{0} =1$.} depends upon the 
ratio $(T/H)$ which measures the ratio between the temperature and the Hubble 
rate when $N_{\mathrm{tot}} \simeq N_{\mathrm{max}}$, i.e.
\begin{equation}
\biggl(\frac{T}{H}\biggr) \ll 144.5 \biggl(\frac{106.75}{g_{\rho}}\biggr)^{1/4} 
\biggl(\frac{2.43\times 10^{-9}}{{\mathcal A}_{{\mathcal R}}}\biggr)^{1/4}
\biggl(\frac{0.01}{\epsilon}\biggr)^{1/4},
\label{DS18A}
\end{equation}
where $g_{\rho}$ denotes the total number of spin degrees of freedom 
for $T> 200$ GeV.  If $N_{\mathrm{tot}} \gg N_{\mathrm{max}}$, 
$k_{\mathrm{T}}$ will become arguably much smaller than $H_{0}$ as Eqs. (\ref{DS18}) and  (\ref{DS18A}) imply immediately.
 
The uncertainty in the estimates of Eqs. (\ref{DS18}) and (\ref{DS18A}) resides not 
only in the (inevitably) unknown 
value of the total number of e-folds but also in
$N_{\mathrm{max}}$ whose value cannot be precisely assessed even within the 
consistent lore provided by the conventional inflationary 
scenarios.  Indeed, the uncertainty affecting the determination of 
$N_{\mathrm{max}}$ is due to the lack of a specific knowledge 
of the post-inflationary thermal history. Suppose, for instance, that 
right at the end of the inflationary phase, the standard radiation-dominated 
phase starts. In this case we have that 
\begin{equation}
N_{\mathrm{max}}  = 62.2 + \frac{1}{2} \ln{\biggl(\frac{\xi}{10^{-5}}\biggr)} - \ln{\biggl(\frac{h_{0}}{0.7}\biggr)}
+ \frac{1}{4} \ln{\biggl(\frac{h_{0}^2 \, \Omega_{\mathrm{R}0}}{4.15\times 10^{-5}}\biggr)}
\label{DS25B}
\end{equation}
Recalling that, according to the WMAP 7 yr data, the amplitude of the scalar power spectrum at the pivot scale $k_{\mathrm{p}} =0.002\, \mathrm{Mpc}^{-1}$ is given by ${\mathcal A}_{{\mathcal R}} = 2.43 \times 10^{-9}$, 
the estimate of Eq. (\ref{DS25}) becomes (see, e.g. \cite{mgprimer})
\begin{equation}
N_{\mathrm{max}} = 63.6 + \frac{1}{4} \ln{\epsilon}.
\label{DS26}
\end{equation}
The figures given in Eqs. (\ref{DS25}) are just a dim indication 
since $N_{\mathrm{max}}$ can indeed be much larger. 
Suppose, for instance, that right after inflation the Universe expands at a rate 
which is slower than radiation. In this case $N_{\mathrm{max}}$ increases. In particular, if, after inflation, the 
energy density of the plasma is dominated by a stiff source with sound speed 
coinciding with the speed of light we get to the estimate 
\begin{equation}
N_{\mathrm{max}} = 78.3 + \frac{1}{3} \ln{\epsilon},
\label{DS27}
\end{equation}
where it has been assumed that the stiff phase starts right after inflation and stops right before big-bang nucleosynthesis
(see, e.g. \cite{lid,mgs,mgs1,mgs2} and references therein).
By definition $N_{\mathrm{max}}$ is derived by requiring that the present 
size of the Hubble radius is all contained in the event horizon at the onset 
of the inflationary phase. 
Being optimistic we can say that $N_{\mathrm{max}}  = 63 \pm 15$ 
which is anyway a pretty large indetermination.
The indetermination on the  specific values of $N_{\mathrm{tot}}$ and $N_{\mathrm{max}}$ implies 
that, unless $N_{\mathrm{tot}} \simeq N_{\mathrm{max}}$,  the thermal  wavelength 
$k_{\mathrm{T}}^{-1}$ will be much larger than the present value of the Hubble rate. 
Bearing in mind the previous caveats,  using Eq. (\ref{DS17}) and recalling
 Eqs. (\ref{DS15})--(\ref{DS15B}) we shall have that ${\mathcal B}_{\mathrm{s}}(\alpha) = |{\mathcal N}(\alpha)|^2/|{\mathcal D}(\alpha)|^2$ where 
\begin{eqnarray}
&& {\mathcal N}(\alpha) = \sqrt{3} \int d\ln{k} \, k^{6 - 4 \alpha} \int dy\, y^{1 - 2 \alpha} \coth{\biggl(\frac{k\, y}{2 k_{\mathrm{T}}}\biggr)}
\int_{|1 - y|}^{|1 + y|} z^{1 - 2\alpha} \coth{\biggl(\frac{k \, z}{2 k_{\mathrm{T}}}\biggr)}
\label{DS28}\\
&& {\mathcal D}(\alpha)  = \sqrt{2} \int d\ln{k} \,k^{3 - 2\alpha} \,\coth{\biggl(\frac{k\, y}{2 k_{\mathrm{T}}}\biggr)}.
\label{DS29}
\end{eqnarray}
The evaluation of the integrals can be performed with different methods and even numerically. It is instructive, however, to derive an explicit analytic estimate based 
on the observation that $\coth{x}$ can be 
approximated with $1/x$ for $x<1$ and with $1$ for $x>1$. The integrals 
of Eqs. (\ref{DS28}) and (\ref{DS29}) can therefore be evaluated by using this simple 
approximation scheme which can be improved by keeping further terms in the expansion. To lowest order we will then have that 
\begin{equation}
{\mathcal D}(\alpha) = \frac{\sqrt{2} k_{\mathrm{T}}}{(1 - \alpha)} \biggl[ ( 2 k_{\mathrm{T}})^{2 - 2 \alpha} - 
k_{\mathrm{min}}^{2 - 2 \alpha}\biggr] + \frac{\sqrt{2}}{3 - 2\alpha}\biggl[ k_{\mathrm{max}}^{3 - 2 \alpha} - 
(2 k_{\mathrm{T}})^{3 - 2 \alpha}\biggr].
\label{DS30}
\end{equation}
The leading and subleading terms in Eq. (\ref{DS30}) are determined by the hierarchy between 
$k_{\mathrm{min}}$, $H_{0}$ and $k_{\mathrm{T}}$. In particular, if
 $k_{\mathrm{min}} > 2 k_{\mathrm{T}}$ then it will always be true that $k > 2 k_{\mathrm{T}}$ since,
 by definition of $k_{\mathrm{min}}$, $k$ cannot be smaller than $k_{\mathrm{min}}$. But then from Eq. (\ref{DS17}) and (\ref{DS29}) 
 $(2\overline{n}_{k} + 1) \simeq 1$ for all the range of the momenta. This situation happens, in particular, 
 when $N_{\mathrm{tot}} \gg N_{\mathrm{max}}$. In this situation the degree of second-order 
 coherence will reproduce the vacuum case, i.e., for $|k\,r|\ll 1$ and $|k \tau | \ll 1$, $g^{(2)}(r,\tau) \to 3$.
 In the opposite situation $k_{\mathrm{min}} > 2 k_{\mathrm{T}}$ but then the leading term in Eq. 
 (\ref{DS30}) will be the one coming from $k_{\mathrm{min}}$ (which will be assumed 
 to coincide with $H_{0}$ for the purpose of numerical estimates). The rationale for the 
 latter statement stems from the value of $(3 - 2\alpha)$ which is always negative in the case of conventional 
 slow-roll dynamics. The same kind of considerations can be used for the estimate 
 of ${\mathcal N}(\alpha)$ which can also be written, after integration over $z$, as
 \begin{eqnarray}
{\mathcal N}(\alpha) &=& \int dk\,\, k^{ 5 - 4\alpha} \,\, \int dy y^{ 1 - 2 \alpha} \coth{\biggl(\frac{k}{ 2 k_{\mathrm{T}}}\biggr)}
\biggl\{ \frac{( 2 k_{\mathrm{T}}/k)^{2 (1 -\alpha)}}{( 1 - 2\alpha) (2 - 2\alpha)} 
\nonumber\\
&+& 
\frac{| 1 + y|^{2( 1 -\alpha)}}{2(1 -\alpha)} - \biggl(\frac{2 k_{\mathrm{T}}}{k} \biggr) \frac{|1 - y|^{(1- 2\alpha)}}{(1  - 
2\alpha)} \biggr\}.
\label{DS31}
\end{eqnarray}
Using Eq. (\ref{DS6}) expressing the relation between the indices $\alpha$ (and $\beta$) and the tensor (and scalar) 
spectral indices, the final result for the degree of 
second-order coherence can be written, in the limit $kr\to 0$ and $k\tau\to 0$ 
\begin{eqnarray}
g^{(2)}(r,\tau)  &=& \frac{3}{2}\biggl[1 - \frac{( n_{\mathrm{t}} -2) (n_{\mathrm{t}} + 2) (n_{\mathrm{t}} + 8)}{2 r_{\mathrm{t}}}\biggr]
\label{DS32}\\
g^{(2)}(r,\tau)  &=&\frac{3}{2}\biggl[1 - \frac{(3 n_{\mathrm{s}} - 4 n_{\mathrm{t}} -11)( 2 n_{\mathrm{s}} - 
n_{\mathrm{t}} - 4)(n_{\mathrm{t}} +2)}{8 (n_{\mathrm{s}} -1)(n_{\mathrm{t}} - n_{\mathrm{s}} +3)}\biggr],
\label{DS33}
\end{eqnarray}
where Eq. (\ref{DS32}) holds for the gravitons while Eq. (\ref{DS33}) holds in the case 
of the scalar phonons. Both expressions have been obtained by using the relations previously derived in Eq.
(\ref{DS5}) and by demanding, as previously explained, that $N_{\mathrm{tot}} \simeq N_{\mathrm{max}}$.
It is interesting to notice that as long as $r_{\mathrm{t}}$ and $n_{\mathrm{s}} -1$ both diminish the second term 
in the square brackets increases above $1$ and it can happen that $g^{(2)}(r,\tau) < 3$ possibly becoming even negative.

Let us now assume, for a moment,  that the intensity correlations are experimentally 
accessible and that $g^{(2)}$ (i.e. the degree of second-order coherence in the large-scale limit) can be directly measured. The considerations 
developed in the present section suggest three considerations:
\begin{itemize}
\item{} if $g^{(2)} = 3$ the HBT correlations would imply that the 
fluctuations are bunched, super-Poissonian and probably coming 
from a vacuum initial state;
\item{} if $g^{(2)} > 3$ the initial state contained second-order correlations 
which are not characterized by Bose-Einstein statistics but which, nonetheless, 
dominate at large scale; this would probably be a remnant of the initial state and would imply, within the inflationary lore, $N_{\mathrm{tot}} \sim N_{\mathrm{max}} \sim N_{\mathrm{min}}$;
\item{} if $2 < g^{(2)} < 3$ the initial conditions are dominated 
by a thermal ensemble and $N_{\mathrm{tot}} \sim N_{\mathrm{max}}$;
\item{} if $ 1 < g^{(2)} \leq 2$ the intensity correlations are still super Poissonian 
but not squeezed.
\end{itemize}
The possibilities listed above are the result of the preliminary discussion 
reported here and must be  sharpened further. At the same time they illustrate 
how a direct study of the intensity correlations would make the problem of the 
large-scale initial conditions much less elusive.  Finally, recalling the terminology 
introduced at the end of section \ref{sec2}, curvature phonons can even be super-chaotic.

The value of $\overline{g}^{(2)}$ measures, in the 
context of HBT interferometry, the statistical tendency of the phonons to distribute themselves 
in bunches rather than randomly and thus obeying a Poissonian distribution. If the degree of second-order 
coherence is larger than $1$ the positive correlation between the particles arriving at the HBT detectors is dubbed as super-Poissonian. In the case of the curvature phonons the degree of second-order coherence can even be 
larger than $2$. In this case it is natural to talk about the possibility that the statistics is super-chaotic 
since the case $\overline{g}^{(2)} =2$ corresponds to the statistics of chaotic (i.e. white) light. 
\begin{figure}[!ht]
\centering
\includegraphics[height=6cm]{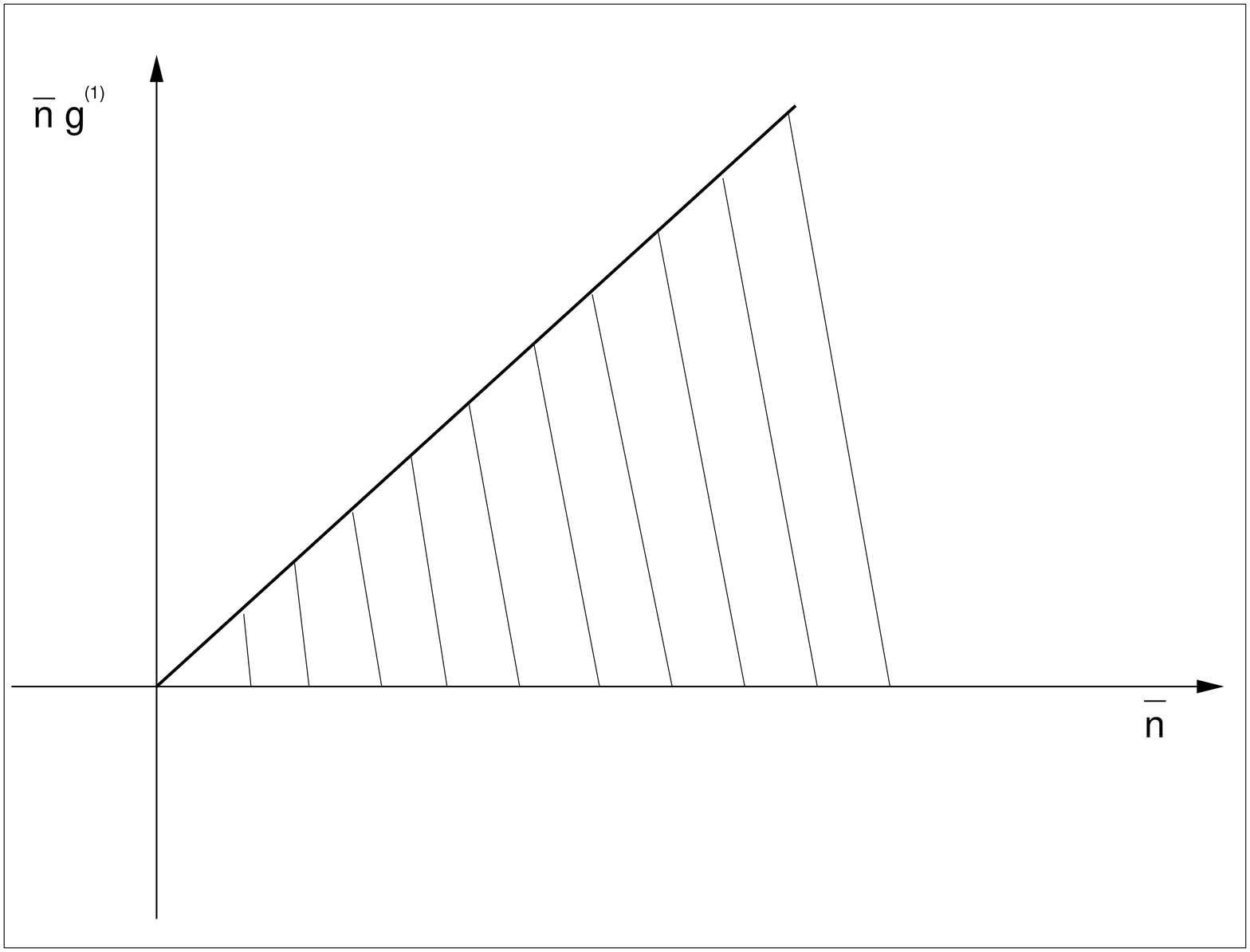}
\includegraphics[height=6cm]{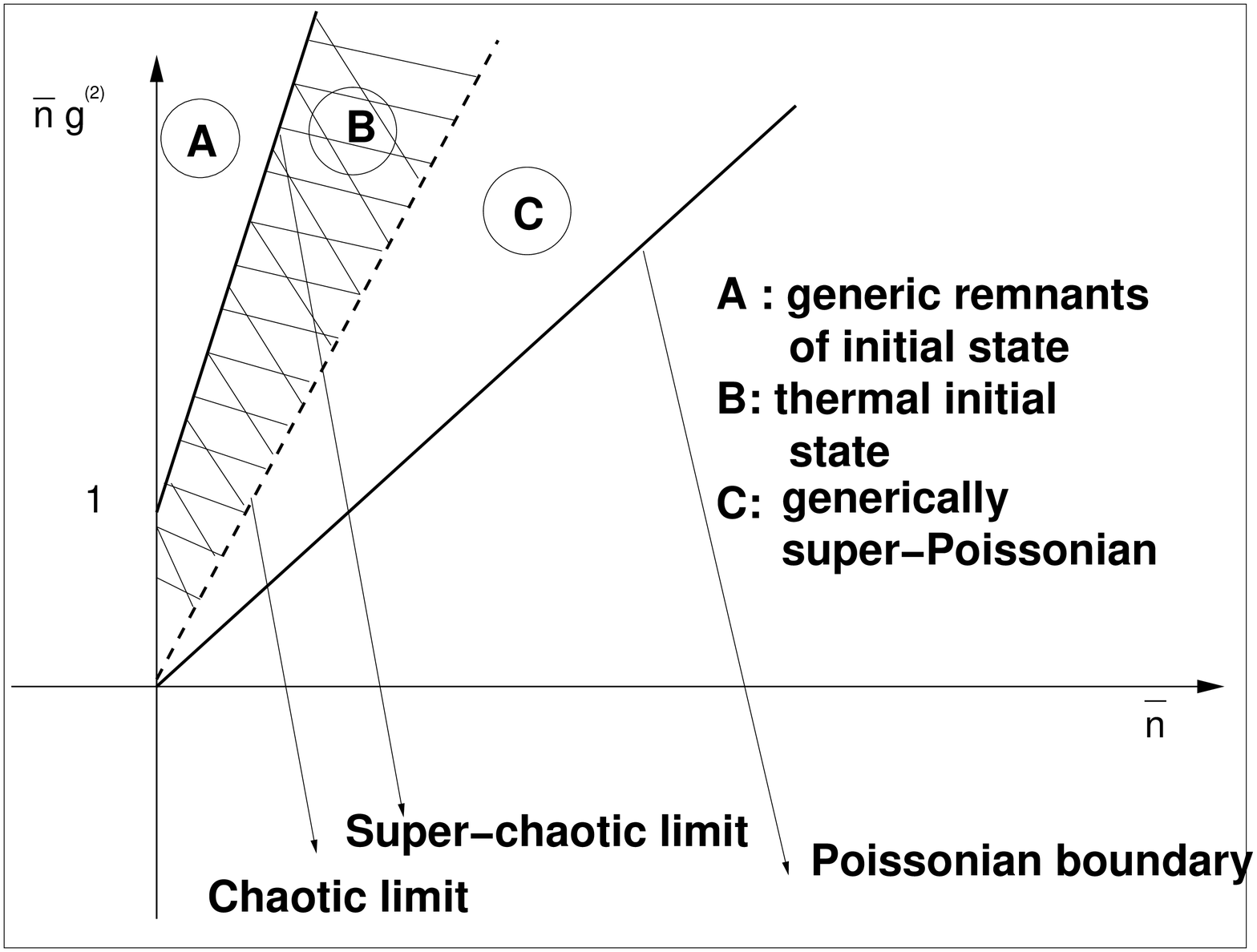}
\caption[a]{The degrees of first-order and second-order coherence are reported in the case of the scalar phonons.}
\label{figure2}      
\end{figure}
The bounds summarized in the previous paragraph are illustrated in Fig. \ref{figure2} where, in the left and right 
plots, the degrees of first-order and second-order coherence are reported. The regions $A$, $B$ and $C$ correspond, respectively, 
to a super-chaotic state, to a squeezed thermal state and to a super-Poissonian statistical ensemble not necessarily related to quantum mechanics.

\renewcommand{\theequation}{7.\arabic{equation}}
\setcounter{equation}{0}
\section{HBT temperature and polarization correlations}
\label{sec7}
Some preliminary considerations will now be developed with the 
purpose of suggesting that HBT correlations could be directly assessed by studying higher order temperature and polarization correlations in the limit of large angular scales. Consider, to begin with, the brightness perturbations which fully describe the temperature and polarization 
anisotropies in the $\Lambda$CDM model neglecting, for simplicity, the tensor modes\footnote{The same 
considerations developed in this section can be however extended to the case of the tensors by 
considering higher-order correlations of the B-mode polarization, if and when they will be measured.} which 
are indeed absent in the vanilla $\Lambda$CDM:
\begin{equation}
\Delta_{\mathrm{I}}(\hat{n},\tau) = \frac{1}{(2\pi)^{3/2}} \int d^{3} k \, \Delta_{\mathrm{I}}(k,\mu,\tau), \qquad 
\Delta_{\mathrm{P}}(\hat{n},\tau) = \frac{1}{(2\pi)^{3/2}} \int d^{3} k \, \Delta_{\mathrm{P}}(k,\mu,\tau);
\label{TP1}
\end{equation}
as usual, $\mu = \hat{k}\cdot \hat{n}$ denotes the projection of the Fourier mode on the direction of propagation 
of the CMB photon\footnote{The angular 
variables $\mu = \hat{k}\cdot \hat{n}$ defined and used in the present section have no relation with $\mu(\vec{x},\tau)$ used and defined in section \ref{sec3} at 
Eq. (\ref{EQQ8});
similarly the angular variable $\varphi$ of Eq. (\ref{TP2}) is not related to the 
inflaton background field.}; notice, furthermore, that $\Delta_{\mathrm{P}}(\hat{n},\tau) = \Delta_{\mathrm{Q}}(\hat{n},\tau)$. 
Since $\Delta_{\pm}(\hat{n},\tau) = \Delta_{\mathrm{Q}}(\hat{n},\tau) \pm i 
 \,\Delta_{\mathrm{U}}(\hat{n}\tau)$, it transforms as a spin $\pm$2 for rotations 
around a plane orthogonal to the direction of propagation of the radiation. The three-dimensional rotations and the rotations on the tangent plane of the sphere at a given point combine to give a $O(4)$ symmetry group \cite{sud}. Generalized ladder operators raising (or lowering) the spin weight of a given function can then be defined as \cite{sud,zalda}:
\begin{equation}
K_{\pm}^{\mathrm{s}}(\hat{n})= - (\sin{\vartheta})^{\pm\mathrm{s}}\biggl[ \partial_{\vartheta} \pm
\frac{i}{\sin{\vartheta}} \partial_{\varphi}\biggr] (\sin{\vartheta})^{\mp\mathrm{s}},\qquad \hat{n} = (\vartheta,\, \varphi).
\label{TP2}
\end{equation}
In real space the E-mode and the B-mode polarization will have spin weight $s=0$
\begin{eqnarray}
&& \Delta_{\mathrm{E}}(\hat{n},\tau) = - \frac{1}{2} \{ K_{-}^{(1)}(\hat{n})[K_{-}^{(2)}(\hat{n})
\Delta_{+}(\hat{n},\tau)] +  K_{+}^{(-1)}(\hat{n})[K_{+}^{(-2)}(\hat{n}) \Delta_{-}(\hat{n},\tau)]\},
\label{TP3}\\
&&  \Delta_{\mathrm{B}}(\hat{n},\tau) =  \frac{i}{2} \{ K_{-}^{(1)}(\hat{n})[K_{-}^{(2)}(\hat{n})
\Delta_{+}(\hat{n},\tau)] -  K_{+}^{(-1)}(\hat{n})[K_{+}^{(-2)}(\hat{n})\Delta_{-}(\hat{n},\tau)]\}. 
\label{TP4}
\end{eqnarray}
and will therefore be invariant under rotations around $\hat{n}$ exactly as $\Delta_{\mathrm{I}}(\hat{n},\tau)$.
Recalling that $\mu = \cos{\vartheta}$ the derivatives with respect to $\vartheta$ can be traded for 
derivatives with respect to $\mu$ and Eqs. (\ref{TP3}) and (\ref{TP4}) imply, in the $\Lambda$CDM framework and in the 
absence of gravitons,
\begin{equation}
\Delta_{\mathrm{E}}(\hat{n}, \tau) = -  \partial_{\mu}^{2} \{( 1 - \mu^2) 
\Delta_{\mathrm{P}}(\hat{n},\tau)\},\qquad \Delta_{\mathrm{B}}(\hat{n}, \tau) = 0,
\label{TP5}
\end{equation}
where the second equality is easily deduced since the derivatives with respect to $\varphi$ vanish in the 
absence of the tensor contribution of the gravitons to the brightness perturbations. 
Therefore, in the $\Lambda$CDM model we have at our disposal three complementary degrees 
of second-order coherence which can be defined 
\begin{eqnarray}
g^{(2)}_{\mathrm{TT}}(\hat{m},\hat{n},\tau) &=& 
\frac{\langle \Delta_{\mathrm{I}}(\hat{m}, \tau) \, \Delta_{\mathrm{I}}(\hat{m}, \tau)\,
\Delta_{\mathrm{I}}(\hat{n}, \tau) \, \Delta_{\mathrm{I}}(\hat{n}, \tau)\rangle}{\langle  \Delta_{\mathrm{I}}(\hat{m}, \tau) \, \Delta_{\mathrm{I}}(\hat{m}, \tau)\rangle \langle  \Delta_{\mathrm{I}}(\hat{n}, \tau) \, \Delta_{\mathrm{I}}(\hat{n}, \tau) \rangle},
\label{TP6}\\
g^{(2)}_{\mathrm{EE}}(\hat{m},\hat{n},\tau) &=& 
\frac{\langle \Delta_{\mathrm{E}}(\hat{m}, \tau) \, \Delta_{\mathrm{E}}(\hat{m}, \tau)\,
\Delta_{\mathrm{E}}(\hat{n}, \tau) \, \Delta_{\mathrm{E}}(\hat{n}, \tau)\rangle}{\langle  \Delta_{\mathrm{E}}(\hat{m}, \tau) \, \Delta_{\mathrm{E}}(\hat{m}, \tau)\rangle \langle  \Delta_{\mathrm{E}}(\hat{n}, \tau) \, \Delta_{\mathrm{E}}(\hat{n}, \tau) \rangle},
\label{TP7}\\
g^{(2)}_{\mathrm{TE}}(\hat{m},\hat{n},\tau) &=& 
\frac{\langle \Delta_{\mathrm{T}}(\hat{m}, \tau) \, \Delta_{\mathrm{E}}(\hat{m}, \tau)\,
\Delta_{\mathrm{T}}(\hat{n}, \tau) \, \Delta_{\mathrm{E}}(\hat{n}, \tau)\rangle}{\langle  \Delta_{\mathrm{T}}(\hat{m}, \tau) \, \Delta_{\mathrm{E}}(\hat{m}, \tau)\rangle \langle  \Delta_{\mathrm{T}}(\hat{n}, \tau) \, \Delta_{\mathrm{E}}(\hat{n}, \tau) \rangle}.
\label{TP8}
\end{eqnarray}
The degrees of second-order coherence defined in Eqs. (\ref{TP6}), (\ref{TP7}) and (\ref{TP8}) 
can be connected with the degree of second-order coherence introduced in sections \ref{sec5} and \ref{sec6}.
In the gauge defined by Eq. (\ref{EQ1}) the brightness perturbations of Eq. (\ref{TP1}) obey, in Fourier space,
\begin{eqnarray}
&&\partial_{\tau}\Delta_{\mathrm{I}} + (i k \mu + \varepsilon') \Delta_{\mathrm{I}} = \partial_{\tau}\psi - i k \mu \phi
 +  \varepsilon' \biggl[\Delta_{\mathrm{I}0} + \mu v_{\mathrm{b}} + \frac{(1 - 3 \mu^2)}{4}S_{\mathrm{P}}(k,\tau)\biggl],
\label{TP9}\\
&& \partial_{\tau}\Delta_{\mathrm{P}} + (i k \mu + \varepsilon') \Delta_{\mathrm{P}} = \frac{3}{4}(1 - \mu^2) S_{\mathrm{P}}(k,\tau),
\label{TP10}
\end{eqnarray}
where $S_{\mathrm{P}}(k,\tau)$ can be expressed as the sum of the 
quadrupole of the intensity, of the monopole of the polarization and 
of the quadrupole of the  polarization, i.e.,   respectively, $S_{\mathrm{P}}(k,\tau) = 
(\Delta_{\mathrm{I}2} + \Delta_{\mathrm{P}0} + \Delta_{\mathrm{P}2})$; note that, in Eqs. (\ref{TP9}) and (\ref{TP10}), 
$\varepsilon'$ and $\varepsilon(\tau,\tau_{0})$ denote, respectively,  the differential optical depth and the optical depth 
itself
\begin{equation}
\varepsilon'= x_{\mathrm{e}} \tilde{n}_{\mathrm{e}} \,a\,\sigma_{\gamma\mathrm{e}}, \qquad 
\varepsilon(\tau,\tau_{0}) = \int_{\tau_{0}}^{\tau}  x_{\mathrm{e}} \tilde{n}_{\mathrm{e}} \,a\,\sigma_{\gamma\mathrm{e}} d\tau
\label{TP10a}
\end{equation}
and should not be confused with $\epsilon$ defined, in the present paper, as one of the slow-roll parameters 
(see, e.g., Eq. (\ref{DS4})). The line of sight solution of Eqs. (\ref{TP9}) and (\ref{TP10}) can be written, respectively, as 
\begin{eqnarray}
\Delta_{\mathrm{I}}(k, \mu, \tau_{0}) &=& \int_{0}^{\tau_{0}} {\mathcal K}(\tau) \biggl[ \Delta_{\mathrm{I}0} + \phi + \mu v_{\mathrm{b}}  + \frac{(1 - 3 \mu^2)}{4} S_{\mathrm{P}}\biggr] e^{- i \mu x(\tau)}
\nonumber\\
&+& \int_{0}^{\tau_{0}} d\tau e^{- \varepsilon(\tau,\tau_{0})} ( \phi' + \psi') e^{-i \mu x(\tau)} d\tau,
\label{TP11}\\
\Delta_{\mathrm{P}}(k,\mu,\tau_{0}) &=& \frac{3}{4}(1-\mu^2) \int_{0}^{\tau_{0}}{\mathcal K}(\tau) S_{\mathrm{P}}(k,\tau) e^{- i k\mu (\tau- \tau_{0})} d\tau,
\label{TP12}
\end{eqnarray}
where ${\mathcal K}(\tau) = \varepsilon' e^{- \varepsilon(\tau,\tau_{0})}$ is the visibility function 
and $x(\tau) = k (\tau_{0} - \tau)$. The visibility function can be approximated as 
a double Gaussian with two peaks roughly corresponding to the redshifts 
of recombination and reionization, i.e. 
$z_{\mathrm{rec}} \simeq 1088.2\pm 1.2$ and $z_{\mathrm{reion}}= 10.5 \pm 1.2$ according to 
\cite{WMAP7a,WMAP7b}. The semi-analytical parametrizations of the visibility function
(such as the ones of \cite{v1,v2}) are relevant when investigating the degree of second-order coherence 
for angular scales than the degree. In the present analysis the focus will be on the large-angular scales 
corresponding to  typical multipoles $\ell \leq \sqrt{z_{\mathrm{rec}}}$ where the finite width 
of the visibility function is immaterial and the opacity suddenly drops at recombination. This 
implies that the visibility function presents a sharp (i.e. infinitely thin) peak at the recombination time.  Thus  ${\mathcal K}(\tau)$ is proportional to a Dirac delta 
function and $e^{- \varepsilon(\tau,\tau_{0})}$ is proportional to an Heaviside theta function.  Under the latter approximations, 
 Eq. (\ref{TP9}) leads to the well known pair of separated contributions, i.e. the Sachs-Wolfe (SW) and the integrated Sachs-Wolfe  (ISW) contributions:
\begin{eqnarray}
&& \Delta_{\mathrm{I}}(k,\mu,\tau_{0}) = \Delta_{\mathrm{I}}^{(\mathrm{SW})}(k,\mu,\tau_{0})  + \Delta_{\mathrm{I}}^{(\mathrm{ISW})}(k,\mu,\tau_{0}), 
\label{HT14a}\\
&& \Delta_{\mathrm{I}}^{(\mathrm{SW})}(k,\mu,\tau_{0}) = \biggl( - \frac{{\mathcal R}(k,\tau)}{5}\biggr)_{\tau_{\mathrm{rec}}} e^{- i \mu y_{\mathrm{rec}}},
\label{HT14}\\
&& \Delta_{\mathrm{I}}^{(\mathrm{ISW})}(k,\mu,\tau_{0}) = \int_{\tau_{\mathrm{rec}}}^{\tau_{0}} 
(\phi' +\psi') e^{- i \mu x(\tau)} \, d\tau,
\label{HT15}
\end{eqnarray}
where, by definition, $x(\tau_{\mathrm{rec}}) = y_{\mathrm{rec}}$.
Equations (\ref{HT14}) and (\ref{HT15}) can be evaluated within various approximation schemes.  The SW and the ISW contributions can be separately evaluated. In particular the ordinary SW contribution becomes 
\begin{eqnarray}
\Delta_{\mathrm{I}}^{(\mathrm{SW})}(k,\mu,\tau_{0}) &=& - \frac{{\mathcal R}(\vec{k},\tau_{\mathrm{i}})}{5} 
{\mathcal S}(q_{\mathrm{rec}}) e^{- i \mu y_{\mathrm{rec}}}  
\label{TP20A}\\
{\mathcal S}(q) &=& 1 + \frac{4}{3 q} - \frac{16}{3 q^2}
+ \frac{16( \sqrt{y +1} -1)}{3 q^3},
\label{TP20B}
\end{eqnarray}
while the ISW contribution is:
\begin{eqnarray}
&&\Delta_{\mathrm{I}}^{(\mathrm{ISW})}(k,\mu,\tau_{0}) = - 2 {\mathcal R}(\vec{k},\tau_{\mathrm{i}}) \int_{\tau_{\mathrm{rec}}}^{\tau_{0}} \partial_{\tau} {\mathcal T}_{\mathcal R}(\tau)e^{- i \mu x(\tau)} \, d\tau,
\label{TP20C}\\
&& {\mathcal T}_{\mathcal R}(\tau) = 1 - \frac{{\mathcal H}(\tau)}{a^2(\tau)} \int_{0}^{\tau} a^2(\tau')\, d\tau'.
\label{TP20D}
\end{eqnarray}
Both in Eqs. (\ref{TP20A}) and (\ref{TP20C}), ${\mathcal R}(\vec{k},\tau)$ denotes the constant value of curvature perturbations 
at $\tau_{\mathrm{i}} < \tau_{\mathrm{eq}}$. By further approximating the integrand in Eq. (\ref{TP20C}) the whole large-scale 
contribution can be written, for the present purposes, as 
\begin{eqnarray}
\Delta_{\mathrm{I}}(k,\tau_{0}) &=&  {\mathcal R}(\vec{k},\tau_{\mathrm{i}})  e^{ - i \mu y_{\mathrm{rec}}},
\nonumber\\
\overline{{\mathcal S}}(q_{\mathrm{rec}}) &=& - \biggl\{  \frac{{\mathcal S}(q_{\mathrm{rec}})}{5} +  \bigl[\partial_{\tau} {\mathcal T}_{\mathcal R} \bigr]_{q_{\mathrm{rec}}}\biggr\},
\label{TP21}
\end{eqnarray}
where
\begin{equation}
q_{\mathrm{rec}} = \frac{a_{\mathrm{rec}}}{a_{\mathrm{eq}}} = \frac{z_{\mathrm{eq}} + 1}{z_{\mathrm{rec}}+1} = 3.04 \biggl( \frac{h_{0}^2 \Omega_{\mathrm{M}0}}{0.134}\biggr).
\label{APP2}
\end{equation}
Equation (\ref{TP21}) directly relates the curvature perturbations to the brightness perturbations. It then follows that 
the two point function of temperature perturbations bears the mark, up to time-dependent factors, of the 
two-point function defined in the context of first-order coherence effects (see section \ref{sec3}). In particular we have that   
\begin{eqnarray}
&&\langle \hat{\Delta}_{\mathrm{I}}(\hat{m},\tau_{0}) \hat{\Delta}_{\mathrm{I}}(\hat{n},\tau_{0})\rangle = \sum_{\ell} \frac{(2 \ell +1)}{4\pi} C_{\ell}^{\mathrm{TT}} \, P_{\ell}(\hat{m}\cdot\hat{n}),
\label{TP22}\\
&& C_{\ell}^{\mathrm{TT}}  = \frac{\overline{{\mathcal S}}^2(q_{\mathrm{rec}})}{4\pi^2 z^2}
\int d\ln{q} \, q^2 \, ( 2 \overline{n}_{q} + 1) \biggl[ 2 \overline{N}_{q} + 1 - 2 \cos{\alpha_{q}} \sqrt{\overline{N}_{q} (\overline{N}_{q} +1)}\biggr]\, j_{\ell}^2[y_{\mathrm{rec}}].
\label{TP23}
\end{eqnarray}
In a similar fashion the degree of second-order coherence can be estimated as
\begin{equation}
\langle \Delta_{\mathrm{I}}(\hat{m}, \tau) \, \Delta_{\mathrm{I}}(\hat{m}, \tau)\,
\Delta_{\mathrm{I}}(\hat{n}, \tau) \, \Delta_{\mathrm{I}}(\hat{n}, \tau)\rangle =\sum_{\ell} (2 \ell + 1) \biggl[ {\mathcal Z}_{\ell} + {\mathcal Q}_{\ell}\, P_{\ell}(\hat{m}\cdot\hat{n})\biggr].
\label{TP24}
\end{equation}
where 
\begin{eqnarray}
&& {\mathcal Z}_{\ell} = \frac{\overline{{\mathcal S}}^2(q_{\mathrm{rec}})}{ z^4} \int d\ln{q} \biggl[ 
2 {\mathcal G}^{(2)}_{\mathrm{v}}(q,\tau_{0})\, j_{\ell}^2(2 y_{\mathrm{rec}}) + 
{\mathcal G}^{(2)}_{\mathrm{s}}(q,\tau_{0}) j_{\ell}^2(y_{\mathrm{rec}})\biggr],
\label{TP25}\\
&& {\mathcal Q}_{\ell} = \frac{\overline{{\mathcal S}}^2(q_{\mathrm{rec}})}{z^4} \int d\ln{q} \biggl[ 
 {\mathcal G}^{(2)}_{\mathrm{v}}(q,\tau_{0})\, j_{\ell}^2(2 y_{\mathrm{rec}}) + 
2 {\mathcal G}^{(2)}_{\mathrm{s}}(q,\tau_{0}) j_{\ell}^2(y_{\mathrm{rec}})\biggr].
\label{TP26}
\end{eqnarray}
The degree of second-order coherence of Eq. (\ref{TP6}) can therefore be written as 
\begin{equation}
g^{(2)}_{\mathrm{TT}}(\hat{m},\hat{n},\tau) = \frac{\sum_{\ell}(2 \ell + 1) [ {\mathcal Z}_{\ell} + {\mathcal Q}_{\ell}\, P_{\ell}(\hat{m}\cdot\hat{n})]}{\bigl|\sum_{\ell} (2 \ell +1) C_{\ell}^{\mathrm{TT}}\bigr|^2}.
\label{TP27}
\end{equation}
All the inequalities established for the degree of second-order coherence  and all the 
considerations presented before are also applicable to Eq. (\ref{TP27}): in the limit 
$\hat{m}\cdot\hat{n} \to 1$, using the well known identities, $g^{(2)}_{\mathrm{TT}}$ 
 coincides with the result expressed, for instance, by Eq. (\ref{DS14}). In the 
cases $\hat{m}\cdot\hat{n} \neq 1$ a specific angular dependence should be taken into account 
when trying to infer the degree of second-order coherence from the observational data. 
The considerations developed in this last section exclude the presence of the tensor modes 
which can be however included without problems.
\renewcommand{\theequation}{8.\arabic{equation}}
\setcounter{equation}{0}
\section{Concluding remarks}
\label{sec8}
In conventional Hanbury Brown-Twiss interferometry the statistical properties of the source are often part of the experimental setup but the space-time dimensions of the emitters need to be determined. For large-scale curvature perturbations the reverse is true: while the statistical properties of the source are unknown, the gross uniformity of the temperature fluctuations at last scattering implies that curvature perturbations prior to matter-radiation equality had typical wavelengths larger 
than the Hubble radius at the corresponding epoch.

Can we directly scrutinize the statistical properties of the large-scale curvature perturbations without 
positing an excessive number of assumptions on the pre-inflationary expansion and on the post-inflationary 
thermal history? This has been the main question addressed in the present paper.
As a partial and preliminary answer, it has been suggested 
that a useful approach to large-scale curvature perturbations of quantum mechanical origin consists in scrutinizing
their large-scale coherence properties. Using then the analogy with a similar problem arising in quantum 
  optics, various results have been obtained and they can be summarized as follows:
\begin{itemize}
\item{} in the limit of wavelengths larger than the Hubble radius  the degree of first-order coherence 
goes always to $1$ in spite of the correlation properties of the initial state;
\item{} the degree of second-order coherence bears  the mark of the statistical properties 
of the initial state; the curvature phonons 
are bunched and their degree of bunching exceeds the typical value of a chaotic source;
\item{} direct limits on (or specific determinations of) the degree  of second-order coherence from temperature and polarization maps 
 can probe the correlation properties of large-scale gravitational fluctuations;
\item{} 
the degree of second-order coherence does depend, in a computable manner,
 upon the values of the slow-roll parameters, upon the nature of the initial state and upon the duration of the inflationary phase ;
\item{} a set of model-independent limits on the degree of second-order coherence has been 
 derived in the form of a collection of inequalities which can be tested explicitly once the degree 
of second-order coherence is defined in terms of the correlators involving the relevant brightness 
parturbations.
\end{itemize}
On a more technical ground, the tenets of the quantum theory of optical coherence have been carefully 
translated to the quantized treatment of the scalar and tensor normal modes of the geometry. The correct quantum mechanical definition of the degree of second-order coherence has been derived and it has been shown to be equivalent, in the limit of a large number of phonons per Fourier mode, to the standard normal-ordered 
definition customarily employed in Hanbury Brown-Twiss interferometry and based on the 
quantum theory of photoelectric detection. 

Pre-inflationary initial conditions are often assigned (or imposed) 
by combining  theoretical prejudice with loose elements of phenomenological 
consistency such as the suppression or the increase of large-scale power spectra.
Instead of arguing that the theoretical prejudice necessarily selects a specific number of inflationary e-folds, a preferred 
set of initial conditions, a unique pre-inflationary history  it is also worthwhile to take a more modest approach and to ask ourselves wether it is possible to gain informations on the correlation properties of pre-decoupling initial conditions 
by using the logic of Hanbury Brown-Twiss interferometry, i.e. the study of the intensity 
correlations of the scalar and tensor fluctuations of the geometry. The results reported in the present analysis 
represent a first step along this direction.


\newpage

\begin{thebibliography}{99}

\bibitem{WMAP7a} C.~L.~Bennett {\it et al.},  arXiv:1001.4758 [astro-ph.CO]; 
N.~Jarosik {\it et al.},  arXiv:1001.4744 [astro-ph.CO].

\bibitem{WMAP7b} J.~L.~Weiland {\it et al.},  arXiv:1001.4731 [astro-ph.CO]; 
 D.~Larson {\it et al.},  arXiv:1001.4635 [astro-ph.CO]; B.~Gold {\it et al.},  arXiv:1001.4555 [astro-ph.GA].  

\bibitem{ACBAR} C.~L.~Reichardt, P.~A.~R.~Ade, J.~J.~Bock {\it et al.}, Astrophys.\ J.\  {\bf 694}, 1200-1219 (2009).

\bibitem{QUAD} M.~Zemcov {\it et al.}  [QUaD collaboration], Astrophys.\ J.\  {\bf 710}, 1541 (2010); 
M.~L.~Brown {\it et al.}  [QUaD collaboration], Astrophys.\ J.\  {\bf 705}, 978 (2009).

\bibitem{wein} S. Weinberg, {\it Cosmology}, (Oxford University Press, Oxford 2009).

\bibitem{LSS1} W.~J.~Percival, B.~A.~Reid, D.~J.~Eisenstein {\it et al.},  Mon.\ Not.\ Roy.\ Astron.\ Soc.\  {\bf 401}, 2148-2168 (2010).

\bibitem{LSS2} B.~A.~Reid, W.~J.~Percival, D.~J.~Eisenstein {\it et al.}, Mon.\ Not.\ Roy.\ Astron.\ Soc.\  {\bf 404}, 60-85 (2010).

\bibitem{SNN1} R.~Kessler, A.~Becker, D.~Cinabro {\it et al.},  Astrophys.\ J.\ Suppl.\  {\bf 185}, 32-84 (2009).

\bibitem{SNN2}  M.~Hicken, W.~M.~Wood-Vasey, S.~Blondin {\it et al.}, Astrophys.\ J.\  {\bf 700}, 1097-1140 (2009).

\bibitem{mg1}  M.~Giovannini,  Class.\ Quant.\ Grav.\  {\bf 20}, 5455-5473 (2003).

\bibitem{mg1a}  M.~Giovannini, Phys.\ Rev.\  D {\bf 67}, 123512 (2003).

\bibitem{nad1} J.~Valiviita, V.~Muhonen,  Phys.\ Rev.\ Lett.\  {\bf 91}, 131302 (2003).

\bibitem{nad2} K.~Enqvist, H.~Kurki-Suonio, J.~Valiviita,Phys.\ Rev.\  {\bf D65}, 043002 (2002).

\bibitem{nad3} R.~Keskitalo, H.~Kurki-Suonio, V.~Muhonen {\it et al.},  JCAP {\bf 0709}, 008 (2007).

\bibitem{sudarshan} J.  Klauder and E. Sudarshan, {\it Fundamentals of quantum optics} (Benjamin, New York, 1968).

\bibitem{loudon}  R. Loudon, {\it The quantum theory of light} (Clarendon Press, Oxford, 1983).

\bibitem{mandel}  L. Mandel and E. Wolf, {\it Optical coherence and quantum optics}, (Cambridge University Press, Cambridge, 1995).

\bibitem{HBT1}  R. Hanbury Brown and R. Q. Twiss, Nature {\bf 178}, 1046 (1956).

\bibitem{HBT2} R. Hanbury Brown and R. Q. Twiss, Proc. Roy. Soc. (London) {\bf A242}, 300 (1957); Proc. Roy. Soc. (London)  {\bf A243}, 291 (1958).

\bibitem{podgo1}  G.~I.~Kopylov, M.~I.~Podgoretsky,  Sov.\ J.\ Nucl.\ Phys.\  {\bf 15}, 219-223 (1972) [Yad. Fiz. {\bf 15}, 392 (1972)]; 

\bibitem{podgo2}   G.~I.~Kopylov, M.~I.~Podgoretsky, Sov.\ J.\ Nucl.\ Phys.\  {\bf 18},  336 (1973) [Yad. Fiz. {\bf 18}, 656 (1973)].

\bibitem{cocconi} G.~Cocconi,  Phys.\ Lett.\  {\bf B49}, 459 (1974).

\bibitem{rev1} D.~H.~Boal, C.~K.~Gelbke, B.~K.~Jennings,   Rev.\ Mod.\ Phys.\  {\bf 62}, 553-602 (1990).

\bibitem{rev2} G. Baym, Acta Phys.\ Polon.\  B {\bf 29}, 1839-1884 (1998).

\bibitem{glauber1} R.~J.~Glauber,  Phys.\ Rev.\ Lett.\  {\bf 10}, 84 (1963).

\bibitem{glauber2} R.~J.~Glauber, Phys.\ Rev.\  {\bf 130}, 2529 (1963);  Phys.\ Rev.\  {\bf 131}, 2766 (1963).

\bibitem{mg2} M.~Gasperini, M.~Giovannini, G.~Veneziano,  Phys.\ Rev.\  {\bf D48}, 439-443 (1993).

\bibitem{mg3}   V.~Bozza, M.~Giovannini, G.~Veneziano,  JCAP {\bf 0305}, 001 (2003).

\bibitem{mg3a} K.~Bhattacharya, S.~Mohanty and R.~Rangarajan, Phys.\ Rev.\ Lett.\  {\bf 96}, 121302 (2006).

\bibitem{mg3b} K.~Bhattacharya, S.~Mohanty and A.~Nautiyal,  Phys.\ Rev.\ Lett.\  {\bf 97}, 251301 (2006).

\bibitem{mollow} B. L. Mollow and R. J. Glauber, Phys. Rev. {\bf 160}, 1076 (1967); {\it ibid}. {\bf 160}, 1097 (1967).

\bibitem{perelomov}  A. Perelomov, {\it Generalized coherent states and their applications}, (Springer-Verlag, Berlin, 1986).

\bibitem{stoler} D.~Stoler,  {\it Phys.\ Rev.\  D} {\bf 1}, 3217 (1970); {\it Phys.\ Rev.\  D} {\bf 4}, 1925 (1971).

\bibitem{yuen} H.~P.~Yuen, Phys.\ Rev.\  {\bf A13}, 2226 (1976).

\bibitem{hollenhorst} J. N. Hollenhorst, Phys. Rev. D. {\bf 19}, 1669 (1979). 

\bibitem{revsq1} B. L. Shumaker, Phys. Rep. {\bf 135}, 317 (1986).

\bibitem{revsq2} J. Grochmalicki and M. Lewenstein, Phys. Rep. {\bf 208}, 189 (1991).

\bibitem{mgprimer} M. Giovannini, {\it A primer on the physics of the Cosmic Microwave Background},  (World Scientific, Singapore 2008).

\bibitem{lid}  A.~R. Liddle and S.~M. Leach,   Phys.\ Rev.\  {\bf D68}, 103503 (2003).

\bibitem{fano} U.~Fano,  Rev.\ Mod.\ Phys.\  {\bf 29}, 74-93 (1957).

\bibitem{karlin} S. Karlin, {\it A first course in stochastic processes}, (Academic Press, New York, 1966).

\bibitem{knight} M.~S.~Kim, F.~A.~M.~de Oliveira, P.~L.~Knight,  Phys.\ Rev.\  {\bf A40}, 2494-2503 (1989).

\bibitem{bard1} J. Bardeen, Phys. Rev. {\bf D22}, 1882 (1980).

\bibitem{bard2} J. Bardeen, P. Steinhardt, and M. Turner, Phys. Rev. {\bf D28}, 679 (1983).

\bibitem{bard3} R.~H.~Brandenberger, R.~Kahn and W.~H.~Press,  Phys.\ Rev.\  {\bf D28}, 1809 (1983).

\bibitem{fordp} L. H. Ford and L. Parker, Phys. Rev. {\bf D16}, 1601 (1977).

\bibitem{sakharov} A. D. Sakharov, Sov. Phys. JETP {\bf 22}, 241 (1966) [Zh. Exp. Teor. Fiz. {\bf 49}, 345 (1965)].

\bibitem{luk} V.~N.~Lukash,  Sov.\ Phys.\ JETP {\bf 52}, 807-814 (1980) [Zh. Eksp. Teor. Fiz. {\bf 79}, 1601 (1980)].

\bibitem{KS}  H.~Kodama, M.~Sasaki,  Prog.\ Theor.\ Phys.\ Suppl.\  {\bf 78}, 1-166 (1984); M. Sasaki, Prog. Teor. Phys. {\bf 76}, 1036 (1986). 

\bibitem{chibisov} G.~V.~Chibisov, V.~F.~Mukhanov,  Mon.\ Not.\ Roy.\ Astron.\ Soc.\  {\bf 200}, 535-550 (1982); V.~F.~Mukhanov,  Sov.\ Phys.\ JETP {\bf 67}, 1297-1302 (1988)  [Zh. Eksp. Teor. Fiz. {\bf 94}, 1 (1988)].

\bibitem{strokov}  V.~Strokov,  Astron.\ Rep.\  {\bf 51}, 431-434 (2007).

\bibitem{solomon} A. I. Solomon J.\ Math.\ Phys.\  {\bf 12}, 390 (1971).

\bibitem{mg4} M.~Giovannini,  Phys.\ Lett.\  {\bf B691}, 274 (2010).

\bibitem{DT}  S.~Deser and C.~Teitelboim,  Phys.\ Rev.\  {\bf D13}, 1592 (1976).

\bibitem{deser} S. Deser,  J. Phys. {\bf A15},  1053 (1982).

\bibitem{mg5} M.~Giovannini,  JCAP {\bf 1004}, 003 (2010).

\bibitem{abr} M. Abramowitz and I. A. Stegun, {\it Handbook of Mathematical Functions} (Dover, New York, 1972).

\bibitem{tric}  A. Erdelyi, W. Magnus, F. Obehettinger, and F. Tricomi,  {\it Higher Trascendental Functions} (Mc Graw-Hill, New York, 1953).  

\bibitem{loudon2} R. Loudon and P. L. Knight, J. Mod. Optics {\bf 34}, 709 (1987).

\bibitem{stenholm} S. Stenholm, Physica Scripta {\bf T12}, 56 (1986).

\bibitem{ford} S.~del Campo, L.~H.~Ford,   Phys.\ Rev.\  {\bf D38}, 3657 (1988).

\bibitem{kapusta} J. Kapusta, {\it Finite-temperature field theory}, (Cambridge University Press, Cambridge 1989).

\bibitem{matsuo} K. Matsuo, Phys. Rev. {\bf A41}, 519 (1990).

\bibitem{mgs} M.~Giovannini, Phys.\ Rev.\  {\bf D60}, 123511 (1999).

\bibitem{mgs1} M.~Giovannini,  Phys.\ Lett.\  {\bf B668}, 44-50 (2008).

\bibitem{mgs2}  M.~Giovannini,   Class.\ Quant.\ Grav.\  {\bf 26}, 045004 (2009).

\bibitem{sud} J.~N.~Goldberg {\it et al.}, J.\ Math.\ Phys.\  {\bf 8}, 2155 (1967).

\bibitem{zalda}  M.~Zaldarriaga and U.~Seljak,  Phys.\ Rev.\  {\bf D55}, 1830 (1997).

  \bibitem{v1} R.~A.~Sunyaev and Y.~B.~Zeldovich,  Astrophys.\ Space Sci.\  {\bf 7}, 3 (1970);
   B. Jones and R. Wyse, Astron. Astrophys. {\bf 149}, 144 (1985).

\bibitem{v2} P. Naselsky and I. Novikov, Astrophys. J. {\bf 413}, 14 (1993); 
H. Jorgensen, E. Kotok, P. Naselsky, and I Novikov, Astron. Astrophys. {\bf 294}, 639 (1995).

 \bibitem{WMAP3a} D.~N.~Spergel {\it et al.} [ WMAP Collaboration ],  Astrophys.\ J.\ Suppl.\  {\bf 170}, 377 (2007).

 \bibitem{WMAP3b}  L.~Page {\it et al.} [ WMAP Collaboration ],  Astrophys.\ J.\ Suppl.\  {\bf 170}, 335 (2007). 

\end{thebibliography}
\end{document}